\documentclass[12pt]{article}
\usepackage[top=36truemm,bottom=36truemm,left=32truemm,right=32truemm]{geometry}

\usepackage{amsmath, amsthm, amssymb}
\usepackage{mathrsfs}
\usepackage[dvipdfmx]{}
\usepackage{bbm}
\usepackage{latexsym}
\usepackage{verbatim}
\usepackage{setspace}
\usepackage{hyperref}
\usepackage{color}
\usepackage{enumerate}
\usepackage[pdftex]{graphicx}
\usepackage[hang,small,bf]{caption}
\usepackage[subrefformat=parens]{subcaption}
\usepackage{here}

\allowdisplaybreaks

\newtheorem{assumption}{Assumption} [section]

\newtheorem{lemma}{Lemma}[section]

\newtheorem{theorem}[lemma]{Theorem}

\newcommand{\QED}{\rightline{$\Box$}}

\newcommand{\indep}{\perp \!\!\! \perp}
\newcommand{\noindep}{\not\!\perp\!\!\!\perp }

\usepackage[round]{natbib}

\begin{document}
{
\singlespacing

\title{Treatment Effects with Multidimensional Unobserved Heterogeneity: \\ Identification of the Marginal Treatment Effect
\thanks{I am grateful to my advisor, Katsumi Shimotsu, for his continuous guidance and support. I am also thankful to Mark Henry, Hidehiko Ichimura, Sokbae Lee, Ryo Okui, Bernard Salani\'e, Edward Vytlacil, and Takahide Yanagi for their insightful comments. Further, I benefited from comments of the seminar participants at The University of Tokyo and Tohoku University.
The author gratefully acknowledges the
support of JSPS KAKENHI Grant Number JP	23KJ0713 and the IAAE travel grant for the 2023 IAAE Conference in Oslo, Norway.
}
\author{Toshiki Tsuda\thanks{Toshiki Tsuda, Department of Economics, Yale University. Email: toshiki.tsuda@yale.edu
} \\
Yale University}
\date{\today}
}
\maketitle
}

\begin{abstract}

This paper establishes sufficient conditions for the identification of the marginal treatment effects with multivalued treatments. Our model is based on a multinomial choice model with utility maximization. Our MTE generalizes the MTE defined in \citet{HV05ecta} in binary treatment models. As in the binary case, we can interpret the MTE as the treatment effect for persons who are indifferent between two treatments at a particular level. Our MTE enables one to obtain the treatment effects of those with specific preference orders over the choice set. Further, our results can identify other parameters such as the marginal distribution of potential outcomes. 

\end{abstract}



\newpage
\section{Introduction}

Assessing heterogeneity in treatment effects is important for precise treatment evaluation. The marginal treatment effect (MTE)  provides rich information on heterogeneity across economic agents regarding their observed and unobserved characteristics. Further, once the MTE is estimated, researchers can obtain other treatment effects, such as the average treatment effect (ATE), the average treatment effect on the treated (ATT), and local ATE (LATE). 

In this paper, we consider the multivalued treatments. While the multivalued treatments complicate the identification of treatment effects, they are often used in many applications. For example, vocational programs provide various types of training to participants, and college choice involves numerous dimensions to respond to varied incentives. The literature has developed treatment effects with multivalued treatments, such as LATE \citep{AI95JASA}, MTE \citep{HRV06RES,HRV08AES, HV07HE71, HP18ecta, LS18ecta} and instrumental variable quantile regression \citep{F20ar}.

For the binary treatment model, \citet{HV99NAS} establish the local instrumental variable (LIV) framework to identify MTE. They assume individuals decide on their choices based on the generalized Roy model that is separable in terms of observed and unobserved variables. \citet{V02ecta} shows that the separable threshold-crossing model in the LIV approach plays the same role as the monotonicity assumption for identifying LATE \citep{IA94ecta}.




For the identification of MTE with multivalued treatments, we examine the multiple discrete choice model based on utility maximization. In this model, the value of each treatment is the sum of an observed term and an unobserved term that represents unobserved heterogeneity. This model is a generalization of the multiple logit model and has been extensively studied in economics since the seminal work of \citet{M74B}. In theoretical research, \citet{M93JE} establishes sufficient conditions for the nonparametric identification of the discrete choice model. In applied research, \citet{D02ECTA} employs this model to study the effect of self-selected migration on returns to college. \citet{KW16QJE} use the discrete multiple-choice model as a self-selection model to analyze the Head Start program’s cost-effectiveness.




We identify the MTE with multidimensional unobserved heterogeneity, which enables us to evaluate treatment effects from multiple perspectives. For instance, we consider three valued treatments and set treatment 0 as the baseline. In this case, the model contains two-dimensional heterogeneity that consists of willingness to take treatment 1 and willingness to take treatment 2 against treatment 0. When we condition the MTE on a high value of the former heterogeneity and a low value of the latter heterogeneity, our identification result reveals the causal effects of those with the preference order treatment 1, treatment 0, and treatment 2 from top to bottom. 


A comparison of our MTE with the MTE with a binary treatment reveals several intriguing similarities and discrepancies. As a similarity, our identified MTE with multivalued treatments has multidimensional heterogeneity whose each element follows a uniform distribution on $(0,1)$ while \citet{HV05ecta} define the MTE with binary treatment conditional on unobserved heterogeneity that also uniformly ranges from 0 to 1.  In this sense, our MTE generalizes the MTE in the binary treatment case defined by \citet{HV05ecta} to the multivalued treatment case. Additionally, as in the binary case, we can interpret the MTE as the treatment effect for persons indifferent between treatment 1 and 0 and treatment 2 and 0 at a specific level.  On the other hand, two MTEs have different relationships between treatments and preference order.  In the binary treatment case, the MTE corresponds to marginal changes in the treatment choice because an individual's preference order exactly maps to his choice. However, in the multivalued treatment case, marginal changes in preference order do not necessarily correspond to the changes in treatments. Therefore, our MTE with multivalued treatments represents marginal changes in preferences over the choice set.





The main challenge of the identification is that the model properties prevent us from obtaining several marginal changes in treatments.  In the binary treatment case,  we set a threshold for selecting the treatment and take a derivative with respect to the threshold.  This procedure identifies the MTE because the derivative with respect to the threshold exactly expresses a marginal change in the treatment. In the case of multivalued treatments, we set one specific treatment as the baseline and construct the multiple-choice model by comparing the other treatments with the baseline. For the identification, we set thresholds of the other treatments compared with the baseline and take derivatives with respect to these thresholds. In the case of the baseline treatment, this procedure identifies the conditional expectation because the derivative with respect to each threshold exactly expresses a marginal change in each treatment against the baseline. In the case of the other treatments, changing thresholds has indirect effects on all the other treatments and we cannot identify conditional expectations of those treatments by simply taking derivatives with respect to thresholds.

We solve indirect effects by focusing on an area of each treatment that thresholds have only a direct effect. In this model, each treatment has one threshold that has both indirect and direct effects on that treatment. We remove the indirect effects of the threshold by an ingenuous transformation that enables us to substitute those indirect effects with the marginal changes in other thresholds. By removing indirect effects with those substitutes, we can extract the direct effect from the marginal change in the threshold and identify conditional expectations of all the treatments from the multiple discrete choice model.

By identifying conditional expectations of treatments, we can obtain several treatment effects, including the MTE. Because our result identifies conditional expectations of each treatment given unobserved heterogeneity, we can obtain the MTE with multivalued treatments by taking their differences. Further, we can also obtain the marginal distribution of potential outcomes, which leads to identifying quantile treatment effects given multidimensional unobserved heterogeneity.




We also establish a sufficient condition for identifying thresholds. In the case of multivalued treatments, the connection between thresholds and propensity scores is unclear, even though the propensity score is equal to the threshold in the binary case. We assume the existence of at least one instrument that significantly and negatively affects only one treatment. This assumption enables us to identify the thresholds and is also used by \citet{LS18ecta} for the identification of thresholds.



In the existing literature on the identification of the MTE with multivalued treatments, \citet{LS18ecta} investigate the identification of conditional expectations given unobserved variables based on multinomial choice models characterized by a combination of separable threshold-crossing rules. They assume the existence of continuous instruments and identify several causal effects with identified thresholds. Our result complements the applicability of their main theorem by introducing a novel identification strategy.

\citet{HV07HE71} and \citet{HRV08AES} expand the LIV approach to a model with multivalued treatments generated by a general unordered choice model. They study identification conditions of several types of treatment effects including the marginal treatment effect of one specified choice versus another choice. They achieve the identification of the MTE by using an identification-at-infinity type argument. Our identification strategy does not depend on the large support assumption. 

With the introduction of new treatment effects for multivalued treatments, \citet{mountjoy2022community} studies the effect of enrollment in 2-year community colleges on upward mobility, such as years of education and future income. Because the main focus of his paper is the effect of policy changes for 2-year entry on the outcome of  2-year colleges, he defines new treatment effects with respect to the marginal change of the instrument pertained to 2-year entry.  Our MTEs are based on unobserved heterogeneities that correspond to marginal changes in preferences between treatments and can express his treatment effects.




 The remainder of this paper is organized as follows: Section 2 proposes the basic settings and notation used in this study. We construct the model through comparisons between treatments. In Section 3, we explain our MTE with multivalued treatments through figures. We highlight similarities and differences between our MTEs and the MTE in the binary case. After we show the identification of the MTE, we add detailed explanations of our identification strategies. Section 4 establishes sufficient conditions for nonparametric identification of the thresholds. We relate our contributions to the literature in Section 5.  In Sections 2--4, we consider the case when the number of treatments is three. Section 6 discusses the general case and identifies the MTE. Section 7  concludes. Proofs of the main results and some auxiliary results are collected in Appendix A. Appendix B provides an economic intuition of the assumption newly imposed in Section 6.

\textbf{Notation.} Let $: = $ denote ``equals by definition,'' and let a.s. denote ``almost surely.'' Let $\mathbbm{1}\{\cdot\}$ denote the indicator function. For random variables $X$ and $Z$, $f_{X}(\cdot)$ denotes the probability density function of $X$. $F_{X|Z}(\cdot)$ and $Q_{X|Z}(\cdot)$ denotes the distribution and quantile functions of $X$ given $Z$, respectively. 

\section{Model}\label{Model and Basic Assumptions}

Let $\mathcal{K}$ denote the set of treatments and assume the set comprising of $K(=|\mathcal{K}|)$ elements. Let $\{Y_{k}:k\in \mathcal{K}\}$ be a potential outcome. $D_{k}$ takes the value one if the agent takes treatment $k$. The observed outcome and treatment are expressed as $D=\sum_{k=0}^{K-1} kD_{k}$ and $Y=\sum_{k=0}^{K-1}D_{k}Y_{k}$, respectively. The data contains covariates $\mathbf{X}$ and instruments $\mathbf{Z}$. Throughout this article, we condition on the value of $\mathbf{X}$ and suppress it from the notation. Let the support of $Y$ and $\mathbf{Z}$ be $\mathcal{Y}\subset\mathbb{R}$ and $\mathcal{Z}\subset\mathbb{R}^{\dim(\mathbf{Z})}$, respectively. 


Let $\mathbf{Q(Z)}$ denote the vector of functions of the instruments $Q_{i}(\mathbf{Z})$. Let $\mathbf{V}$ be a vector of unobserved continuous random variables.  For some $k\in\mathcal{K}$, define $S_{k}(\mathbf{V},\mathbf{Q(Z)}):=\mathbbm{1}(V_{k}<Q_{k}(\mathbf{Z}))$. $\mathbf{V}$ is a vector of unobserved heterogeneity and $Q_{k}(\mathbf{Z})$ serves as a threshold for each $S_{k}$ when $\mathbf{Z}$ is given. Hence, $S_{k}$ consists of a separable threshold-crossing model as in the generalized Roy model. 

We define MTE as 
\[
E[Y_{k}-Y_{j}|\mathbf{V}]\quad \text{ for $k,j\in\mathcal{K}$}
\]
and analyze sufficient conditions for the identification. As in \citet{HRV06RES,HRV08AES} and \citet{HV07HE71}, we consider the discrete choice model based on utility maximization. This model setting enables us to interpret the above MTE as the treatment effect with unobserved heterogeneity of preferences over the choice set. For details, see Section \ref{interpretation}.


\subsection{Multiple Discrete Choice Model and Basic Assumptions}\label{Model_description}


 For each $k$, we define $R_{k}(\mathbf{Z})$ as an unknown function that maps from $\mathbb{R}^{\dim(\mathbf{Z})}$ to $\mathbb{R}$ and define $U_{k}$ as an unobserved continuous random variable whose support is $\mathbb{R}$. By extending the definition of the treatment variable in the binary treatment model, we formulate the treatment decision as follows:
\begin{equation}\label{model_H}
D_{k}:=\mathbbm{1}\{U_{k}-R_{k}(\mathbf{Z})>\max_{j\neq k}(U_{j}-R_{j}(\mathbf{Z}))\},
\end{equation}
where $\Pr((U_{k}-R_{k}(\mathbf{Z}))=(U_{j}-R_{j}(\mathbf{Z})))=0$ for $j\neq k$.

 Intuitively, by interpreting $U_{k}$ and  $R_{k}(\mathbf{Z})$ as unobserved and observed terms in an agent's utility, this discrete multiple-choice model states that he makes a choice based on utility maximization. From this intuition, we regard model (\ref{model_H}) as a straightforward generalization of the generalized Roy model. 

Model (\ref{model_H}) has been studied extensively in economics since the seminal work of \citet{M74B}. \citet{M93JE} establishes sufficient conditions for the nonparametric identification of utility functions and the joint distribution function of unobserved random terms. The multinomial choice model has also been used in applied research. \citet{D02ECTA} uses this model to study the effect of self-selected migration on the return to college. \citet{KW16QJE} adopt the discrete multiple-choice model as a self-selection model and analyze the Head Start program’s cost-effectiveness in the presence of substitute preschools. \citet{KLM16QJE} examine the effect of types of education on several gains in earnings. They find that the estimated payoffs are consistent with agents choosing fields based on the discrete multiple-choice model.


 For the identification of the MTE, we construct a model through a combination of threshold-crossing models based on model (\ref{model_H}). For simplicity, through Sections 2-4, we examine the three valued treatment case, namely treatment 0, 1, and 2, and we generalize results in Section 6. Without loss of generality, we regard treatment 0 as the baseline and set $U_{0}-R_{0}(\mathbf{Z})=0$ almost surely. We construct a model with three alternatives using two indicator functions. Assume $U_{1}$ and $U_{2}$ are continuously distributed.
Let 
\begin{align*}
Q_{1}(\mathbf{Z})&:=F_{U_{1}}(R_{1}(\mathbf{Z})), \ \ 
&Q_{2}(\mathbf{Z})&:=F_{U_{2}}(R_{2}(\mathbf{Z})), \\
V_{1}&:=F_{U_{1}}(U_{1}), \ \ 
&V_{2}&:=F_{U_{2}}(U_{2}).
\end{align*}
Set
\begin{gather*}
S_{1}=\mathbbm{1}\{V_{1}<Q_{1}(\mathbf{Z})\},\ \ S_{2}=\mathbbm{1}\{V_{2}<Q_{2}(\mathbf{Z})\}.
\end{gather*}
Note that
\begin{align*}
V_{1}<Q_{1}(\mathbf{Z})&\Leftrightarrow F_{U_{1}}(U_{1})<F_{U_{1}}(R_{1}(\mathbf{Z})), \\
&\Leftrightarrow U_{1}-{R}_{1}(\mathbf{Z})<0,
\end{align*}
and a similar argument gives
\begin{align*}
V_{2}<Q_{2}(\mathbf{Z})&\Leftrightarrow {U}_{2}-{R}_{2}(\mathbf{Z})<0.
\end{align*}

Two indicator functions, $S_{1}$ and $ S_{2}$, correspond to comparisons of utilities between treatment 0 and 1, and treatment 0 and 2, respectively. From model \eqref{model_H}, individuals take treatment 0 when the utility of treatment 0 is the highest among all the alternatives. Therefore, we obtain $D_{0}=S_{1}S_{2}$.

We introduce an indicator function that compares utilities between treatments 1 and 2. Define 
\begin{equation*}
        S_{3}:=\mathbbm{1}\{V_{1}<F_{U_{1}}(F_{U_{2}}^{-1}(V_{2})-F_{U_{2}}^{-1}(Q_{2}(\mathbf{Z}))+F_{U_{1}}^{-1}(Q_{1}(\mathbf{Z}))) \}.
\end{equation*}
By trivial calculation, we obtain
\begin{equation}\label{treatment_1-2}
  \begin{split} 
 &V_{1}<F_{U_{1}}(F_{U_{2}}^{-1}(V_{2})-F_{U_{2}}^{-1}(Q_{2}(\mathbf{Z}))+F_{U_{1}}^{-1}(Q_{1}(\mathbf{Z}))), \\
    \Leftrightarrow &F_{U_{1}}^{-1}(V_{1})-F_{U_{2}}^{-1}(V_{2})<F_{U_{1}}^{-1}(Q_{1}(\mathbf{Z}))-F_{U_{2}}^{-1}(Q_{2}(\mathbf{Z})),  \\
    \Leftrightarrow & U_{1}-U_{2}<R_{1}(\mathbf{Z})-R_{2}(\mathbf{Z}), \\
    \Leftrightarrow & U_{1}-R_{1}(\mathbf{Z})<U_{2}-R_{2}(\mathbf{Z}). 
  \end{split}   
\end{equation}
Hence, we have $D_{2}=(1-S_{2})S_{3}$ by definition. A similar argument reveals $D_{1}=(1-S_{1})(1-S_{3})$.


\begin{figure}[htbp]
\hspace{-15mm}
\includegraphics[scale=0.45]{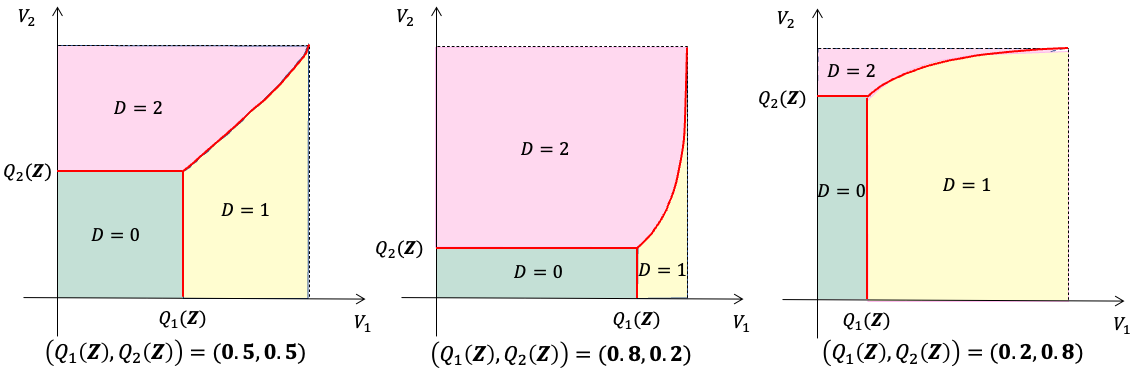}
\caption{This figure shows model (\ref{model_H}) at several values of thresholds when we define $(U_{1},U_{2})\sim N((0,2)^{\top},\Sigma)$ and $\Sigma:=((1.5,3)^{\top},(3,2)^{\top})$.}
\label{figure_model}
\end{figure}

This model is depicted in Figure \ref{figure_model}. In this setting, treatment 0 has the form of the double hurdle model, namely $D_{0}=1$ if and only if $V_{1}<Q_{1}(\mathbf{Z})$ and $V_{2}<Q_{2}(\mathbf{Z})$ as in \citet{LS18ecta}. Even though the double hurdle model essentially expresses the binary treatment case, we successfully specify $D_{1}$ and $D_{2}$ by introducing $S_{3}$ and construct the multiple-choice model based on utility maximization. Hence, our model is not covered by \citet{LS18ecta}. For details, see Section \ref{LS-comaprison}.


We introduce basic assumptions frequently required in the literature on program evaluation.
\begin{assumption}\label{ass_m3}
$\{V_{1}<Q_{1}(\mathbf{Z})\},\{V_{2}<Q_{2}(\mathbf{Z})\}$ and $\{V_{1}<F_{U_{1}}(F_{U_{2}}^{-1}(V_{2})-F_{U_{2}}^{-1}(Q_{2}(\mathbf{Z}))+F_{U_{1}}^{-1}(Q_{1}(\mathbf{Z})))  \}$ are measurable sets.
\end{assumption}

\begin{assumption}[Conditional Independence of Instruments]\label{ass_ind3}
$Y_{0}$, $Y_{1}$, $Y_{2}$ and $\mathbf{V}=(V_{1},V_{2})^{'}$ are jointly independent of $\mathbf{Z}$.
\end{assumption}

\begin{assumption}[Continuously Distributed Unobserved Heterogeneity in the Selection Mechanism]\label{ass_V3}
 The joint distribution of $(U_{1},U_{2})$ is absolutely continuous with respect to the Lebesgue measure on $\mathbb{R}^{2}$. 
\end{assumption}

\begin{assumption}[The Existence of the Moments]\label{ass_mom3}
For $k=0,1,2$, $E[|G(Y_{k})|]<\infty$, where $G$ is a measurable function defined on the support $\mathcal{Y}$ of $Y$, which can be discrete, continuous, or multidimensional.
\end{assumption}

Assumption \ref{ass_m3} ensures the existence of probability of each treatment, that is, $\Pr(D_{k}=1)$. This assumption guarantees that each treatment variable $D_{k}$ is a random variable.  Assumption \ref{ass_ind3} corresponds to the exogeneity of instruments, which plays a vital role in the identification in the literature using instrumental variables. We guarantee the existence of the probability density function of $(U_{1},U_{2})$ by Assumption \ref{ass_V3}. Through the argument of the change of variables, we can also ensure the joint density of $\mathbf{V}$. Assumption \ref{ass_mom3} ensures the existence of moments for each alternative. Otherwise, we cannot define conditional expectations of potential outcomes or identify MTE. Assumptions above often appear in the literature on treatment effects with endogeneity. For instance, Assumptions \ref{ass_m3}, \ref{ass_ind3}, and \ref{ass_V3} correspond to Assumptions 2.1, 2.2, and 3.2 of \citet{LS18ecta}, respectively. Assumption \ref{ass_mom3} generalizes Assumption (A-3) of \citet{HRV08AES}.


\section{Identification}\label{Model and Identification}

\subsection{MTE with Multivalued Treatments}\label{interpretation}


%

In this paper, we study the identification of the following conditional expectations
\begin{equation*}
    E[G(Y_{k})|V_{1}=q_{1}^{*},V_{2}=q_{2}^{*}]\quad \text{for $k\in\{0,1,2\}$ and $(q_{1}^{*},q_{2}^{*})\in(0,1)^{2}$},
\end{equation*}
where we define $(q_{1}^{*},q_{2}^{*})$ as points where we evaluate treatment effects. When we set $G(Y)=Y$ and take the difference between two conditional expectations, we identify the following MTE:
\begin{equation}\label{MTE_T}
E[Y_{k}-Y_{j}|V_{1}=q_{1}^{*},V_{2}=q_{2}^{*}] \quad \text{for $k,j\in\{0,1,2\}$ and $k\neq j$.}
\end{equation}
By definition, each element of $(V_{1},V_{2})$ has a uniform distribution on $(0, 1)$ and  $(q^{*}_{1},q_{2}^{*})$ refer to quantiles of distributions of $U_{1}$ and $U_{2}$, respectively. Hence, $V_{1}$ and $V_{2}$ mean the willingness to choose treatments 1 and 2 compared to treatment 0. For instance, a low value of $V_{1}$ implies an individual is less likely to take treatment 1 than treatment 0. 

MTE \eqref{MTE_T} provides rich information about treatment effects conditioned on individuals' preferences over the choice set. The MTE characterizes preferences among all the alternatives through the values of $(V_{1},V_{2})$. For example, when we identify MTE with a low value of $V_{1}$ and a high value of $V_{2}$, we can interpret this MTE as the average treatment effect in those who are more likely to take treatment 2 and less likely to take treatment 1 compared to treatment 0, i.e., their preferences would be treatment 2, treatment 0 and treatment 1 from top to bottom. 

As another interpretation, MTE \eqref{MTE_T} is the average treatment effect for individuals who would be indifferent between treatment 1 and 0, and treatment 2 and 0 at $(Q_{1}(\mathbf{Z}), Q_{2}(\mathbf{Z}))=(q_{1}^{*}, q_{2}^{*})$. Under Assumption \ref{ass_ind3}, we can illustrate this interpretation in the following equation,
\begin{align*}
&E[Y_{k}-Y_{j}|V_{1}=q_{1}^{*},V_{2}=q_{2}^{*}] \\
=&E[Y_{k}-Y_{j}|V_{1}=q_{1}^{*},V_{2}=q_{2}^{*}, Q_{1}(\mathbf{Z})=q_{1}^{*},Q_{2}(\mathbf{Z})=q_{2}^{*}].
\end{align*}
Note that each unobserved heterogeneity only focuses on comparing two treatments,  treatment 1 and 0, and treatment 2 and 0. Therefore, MTE \eqref{MTE_T} corresponds to the marginal changes in treatment 1 and 0, and treatment 2 and 0.

Comparing MTE (\ref{MTE_T}) with the MTE in the binary treatment case provides useful insights. \citet{HV05ecta} define $D^{*}=1$ as the receipt of the treatment and characterize the decision rule as the generalized Roy model, that is, 
\begin{equation*}
    D^{*}=\mathbbm{1}\{\mu_{D}(\mathbf{Z})-U_{D}\geq 0\},
\end{equation*}
where $\mu_{D}(\mathbf{Z})$ is an unknown function, which maps from  $\mathbb{R}^{\dim(\mathbf{Z})}$ to $\mathbb{R}$, and $U_{D}$ is an unobserved continuous random variable. As a normalization, they innocuously assume that $U_{D}\sim U[0,1]$ and $U_{D}$ is the quantile of the willingness to participate in the treatment. \citet{HV05ecta} then  define MTE with binary treatment as 
\begin{equation}\label{MTE-binary}
    \Delta^{MTE}(u_{D})\equiv E[Y_{1}-Y_{0}|U_{D}=u_{D}],
\end{equation}
where $u_{D}\in (0,1)$. In our definition of MTE with multivalued treatments, $(V_{1},V_{2})$ precisely corresponds to $U_{D}$ in the binary treatment case. Therefore, MTE (\ref{MTE_T}) is a natural generalization of MTE with binary treatment to the multivalued treatment case.

 \citet{HV05ecta} show that treatment effects such as ATE and ATT can be expressed as a function of their MTE. Similarly, in our model, treatment effects such as ATE and ATT can be expressed as a function of our MTE.

MTE \eqref{MTE_T} has a different interpretation from MTE \eqref{MTE-binary} due to the existence of multivalued treatments. In the binary case, whether an individual takes treatment or not precisely corresponds to his preference for its treatment. However, in the multivalued treatment case, preference orders over the choice set have additional information over revealed treatments. For example, if an individual's best treatment is treatment 2, her preference order of the choice set is treatment 2, 0, 1 or treatment 2, 1, 0.  When $Q_{2}(\mathbf{Z})$ marginally changes through $R_{2}(\mathbf{Z})$, this change corresponds to the binary choice between treatment 2 and 0 or treatment 2 and 1, but the change in $R_{2}(\mathbf{Z})$  does not affect the preference between treatment 1 and 0. On the other hand, when $Q_{1}(\mathbf{Z})$ marginally changes and $Q_{2}(\mathbf{Z})$ remains fixed,  her choice may remain in treatment 2 because $Q_{1}(\mathbf{Z})$ only affects the change in preference between treatment 1 and 0. Therefore, marginal changes in $Q_{1}(\mathbf{Z})$ and $Q_{2}(\mathbf{Z})$ correspond to not marginal changes in treatments but marginal changes in preferences between treatment 1 and 0, and treatment 2 and 0, respectively. MTE \eqref{MTE_T} is the treatment effect depicting marginal changes in preferences over the choice set.




\subsection{Illustration}

We illustrate figures of three MTEs, $E[Y_{1}-Y_{0}|\mathbf{V}], E[Y_{2}-Y_{0}|\mathbf{V}]$ and $E[Y_{2}-Y_{1}|\mathbf{V}]$ given $V_2$ is fixed at $0.5$. We depict the MTEs in the following two cases:

 \subsubsection{Case 1: $\mathbf{(V_1,V_2)}$ are not independent of the outcome variable $\mathbf{Y_{k}}$}
 
     \begin{figure}[hbt]
\centering
        \includegraphics[width=0.8\linewidth]{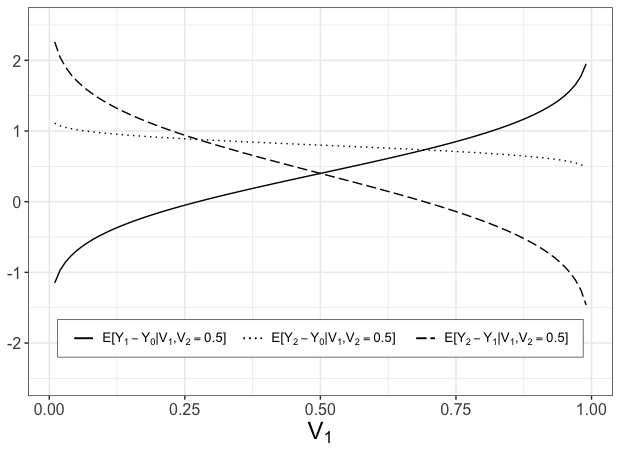}
     \caption{The three MTEs in Case 1. We define $V_{1}$ and $V_{2}$ as $F_{U_{1}}(U_{1})$ and $F_{U_{2}}(U_{2})$, respectively. The model to generate this figure is the following:}
\begin{equation*}
\begin{pmatrix}
Y_0  \\
Y_1 \\
Y_2 \\
U_1 \\
U_2 
\end{pmatrix}
\sim N
\left(
\begin{pmatrix}
0 \\
0.4 \\
0.8 \\
0 \\
0 
\end{pmatrix},
\begin{pmatrix}
1 & -0.2 & -0.2 & -0.2 & -0.2 \\
-0.2 & 1 & 0.2 & 0.5 & 0.2\\
-0.2 & 0.2 & 1 & 0.2 & 0.5\\
-0.2 & 0.5 & 0.2 & 1 & 0.5\\
-0.2 & 0.2 & 0.5 & 0.5 & 1
\end{pmatrix}
\right).
\end{equation*}
\end{figure}

    Case 1 examines three MTEs when $(V_1,V_2)$ is correlated with the outcome variable $Y_{k}$, i.e. 
    \[V_{k}\noindep Y_{\ell}\quad \text{for $k\in\{1,2\}$ and $\ell\in\{0,1,2\}$.}\]
In this case, $E[Y_{1}-Y_{0}|\mathbf{V}]$ increases and $E[Y_{2}-Y_{1}|\mathbf{V}]$ decreases with $V_1$ because people are more likely to choose treatment 1 than treatment 0 at the high value of $V_1$. Even though $V_1$ does not directly affect the difference between treatment 2 and 0, the MTE $E[Y_{2}-Y_{0}|\mathbf{V}]$ decreases slightly with $V_1$, reflecting the combined effect of the decrease in $E[Y_{2}-Y_{1}|\mathbf{V}]$ and the increase in $E[Y_{1}-Y_{0}|\mathbf{V}]$.





\newpage

\subsubsection{Case 2: $\mathbf{V_1}$ is independent of $\mathbf{Y_{0}}$ and $\mathbf{Y_{2}}$ given $\mathbf{V_2}$}

      \begin{figure}[htb]
\centering
        \includegraphics[width=0.8\linewidth]{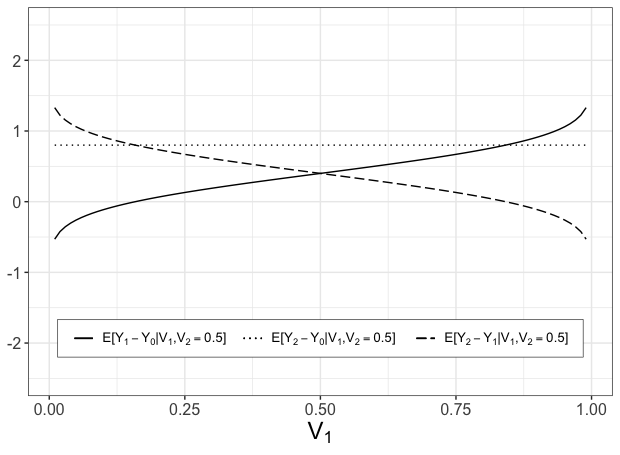}
\caption{The three MTEs in Case 2. We define $V_{1}$ and $V_{2}$ as $F_{U_{1}}(U_{1})$ and $F_{U_{2}}(U_{2})$, respectively. The model to generate this figure is the following:}
\begin{equation*}
\begin{pmatrix}
Y_0  \\
Y_1 \\
Y_2 \\
U_1 \\
U_2 
\end{pmatrix}
\sim N
\left(
\begin{pmatrix}
0 \\
0.4 \\
0.8 \\
0 \\
0 
\end{pmatrix},
\begin{pmatrix}
1 & -0.2 & -0.2 &-0.1 & -0.2 \\
-0.2 & 1 & 0.2 & 0.4 & 0.2\\
-0.2 & 0.2 & 1 & 0.2 & 0.4\\
-0.1 & 0.4 & 0.2 & 1 & 0.5\\
-0.2 & 0.2 & 0.4 & 0.5 & 1
\end{pmatrix}
\right).
\end{equation*}
\end{figure}

    Case 2 analyzes three MTEs when $V_1$ is independent of $Y_{0}$ and $Y_{2}$ given $V_{2}$, i.e. 
    \[V_{1}\indep Y_{k}\quad\text{given $V_2$}\quad  \text{for $k\in\{0,2\}$.}\]
In this case,the MTE $E[Y_{2}-Y_{0}|\mathbf{V}]$ does not depend on $V_1$ and is equal to $E[Y_2-Y_0|V_2=0.5]$ for any $V_{1}\in(0,1)$. Conditional independence of $V_{1}$ implies that the comparison in preference between treatment 1 and 0 does not affect treatment effect $Y_{2}-Y_{0}$. Furthermore, the difference in two other MTEs becomes constant because $E[Y_{2}-Y_{0}|\mathbf{V}]=E[Y_{2}-Y_{1}|\mathbf{V}]+[Y_{1}-Y_{0}|\mathbf{V}]$ holds for any $\mathbf{V}$.

\subsection{Identification Result}

We introduce assumptions to identify the MTE with multivalued treatments. Assumption \ref{new_assumption_dif} is a technical assumption for the proof of the identification, such as continuity and differentiability. 

\begin{assumption}\label{new_assumption_dif}
\
\begin{enumerate}[(1).]
    \item \label{diff_q}
For $k\in\{0,1,2\}$,  $E[G(Y)D_{k}|Q_{1}(\mathbf{Z}), Q_{2}(\mathbf{Z})]$ and $E[D_{k}|Q_{1}(\mathbf{Z}), Q_{2}(\mathbf{Z})]$ are twice differentiable at $(q_{1}^{*}, q_{2}^{*})$.
   \item \label{continuity_v} For $k\in\{0,1,2\}$,  $E[G(Y_{k})|V_{1}, V_{2}]$ and $E[D_{k}|V_{1}, V_{2}]$ are continuous on $(0,1)^{2}$.  
   \item \label{bounded} 
   \begin{enumerate}
   \item \label{bounded_moment}      
   For $k\in\{0,1,2\}$,  $\sup_{(v_{1},v_{2})\in(0,1)^{2}}E[|G(Y_{k})||V_{1}=v_{1},V_{2}=v_{2}]$ is finite.
     \item \label{bounded_conditional} Conditional density functions satisfy the following:
        \begin{equation*}
            \sup_{(u_{1},u_{2})\in\mathbb{R}^{2}}\frac{f_{U_{1},U_{2}}(u_{1},u_{2})}{f_{U_{1}}(u_{1})}<\infty, \quad \sup_{(u_{1},u_{2})\in\mathbb{R}^{2}}\frac{f_{(U_{1},U_{2})}(u_{1},u_{2})}{f_{U_{2}}(u_{2})}<\infty,
        \end{equation*}
   \end{enumerate}
 \end{enumerate}
\end{assumption}
Assumption \ref{new_assumption_dif} \eqref{diff_q} guarantees the existence of derivatives for each conditional expectation. This condition implicitly assumes that the value of $Q_{1}(\mathbf{Z})$ is movable while $Q_{2}(\mathbf{Z})$ is fixed and vice versa. We require Assumption \ref{new_assumption_dif} \eqref{bounded} to exchange differentiation and integration. Assumption \ref{new_assumption_dif} \eqref{bounded_moment} holds when $G(Y_{k})$ is  bounded for each $k\in\{0,1,2\}$.

Conditional on the assumption that $Q_{1}(\mathbf{Z})$ and $Q_{2}(\mathbf{Z})$ are identified, we can identify conditional expectations $E[G(Y_{0})|V_{1}=q_{1}^{*},V_{2}=q_{2}^{*}]$, $E[G(Y_{1})|V_{1}=q_{1}^{*},V_{2}=q_{2}^{*}]$ and $E[G(Y_{2})|V_{1}=q_{1}^{*},V_{2}=q_{2}^{*}]$ by partially differentiating conditional expectations $E[G(Y)D_{k}|Q_{1}(\mathbf{Z})=q_{1},Q_{2}(\mathbf{Z})=q_{2}]$ and $E[D_{k}|Q_{1}(\mathbf{Z})=q_{1},Q_{2}(\mathbf{Z})=q_{2}]$ for $k=0,1,2$.
\begin{theorem}\label{the_iden}
Assume Assumptions \ref{ass_m3} to \ref{new_assumption_dif} hold. Then, conditional expectations of $G(Y_{0}),G(Y_{1})$ and $G(Y_{2})$ are given by
\begin{align*}
\text{(a)} \quad &E[G(Y_{0})|V_{1}=q_{1}^{*},V_{2}=q_{2}^{*}]=\frac{\Delta GD^{*}_{(1,1)}(0)}{\Delta D^{*}_{(1,1)}(0)}, \\
\text{(b)} \quad &E[G(Y_{1})|V_{1}=q_{1}^{*},V_{2}=q_{2}^{*}] \notag  \\
=&-\frac{\Delta GD^{*}_{(1,1)}(1)}{\Delta D^{*}_{(1,1)}(0)}-\frac{\Delta GD^{*}_{(0,1)}(1)[\Delta D^{*}_{(1,0)}(0)+\Delta D^{*}_{(1,0)}(1)]\Delta D^{*}_{(0,2)}(1)}{\Delta D^{*}_{(1,1)}(0)(\Delta D^{*}_{(0,1)}(1))^{2}} \\
&+\frac{\Delta GD^{*}_{(0,2)}(1)[\Delta D^{*}_{(1,0)}(0)+\Delta D^{*}_{(1,0)}(1)]+\Delta GD^{*}_{(0,1)}(1)[\Delta D^{*}_{(1,1)}(0)+\Delta D^{*}_{(1,1)}(1)]}{\Delta D^{*}_{(1,1)}(0)\Delta D_{(0,1)}^{*}(1)}, \notag  \\
\text{(c)} \quad &E[G(Y_{2})|V_{1}=q_{1}^{*},V_{2}=q_{2}^{*}] \\
=&-\frac{\Delta GD^{*}_{(1,1)}(2)}{\Delta D^{*}_{(1,1)}(0)}-\frac{\Delta GD^{*}_{(1,0)}(2)[\Delta D^{*}_{(0,1)}(0)+\Delta D^{*}_{(0,1)}(2)]\Delta D^{*}_{(2,0)}(2)}{\Delta D^{*}_{(1,1)}(0)(\Delta D^{*}_{(1,0)}(2))^{2}} \\
&+\frac{\Delta GD^{*}_{(2,0)}(2)[\Delta D^{*}_{(0,1)}(0)+\Delta D^{*}_{(0,1)}(2)]+\Delta GD^{*}_{(1,0)}(2)[\Delta D^{*}_{(1,1)}(0)+\Delta D^{*}_{(1,1)}(2)]}{\Delta D^{*}_{(1,1)}(0)\Delta D^{*}_{(1,0)}(2)}, 
\end{align*}
where we define
\begin{align*}
    \Delta GD^{*}_{(\ell,m)}(k)&:=\left.\frac{\partial^{(\ell+m)} E[G(Y)D_{k}|Q_{1}(\mathbf{Z}),Q_{2}(\mathbf{Z})]}{\partial^{\ell} Q_{1}(\mathbf{Z})\partial^{m} Q_{2}(\mathbf{Z})}\right|_{(Q_{1}(\mathbf{Z}),Q_{2}(\mathbf{Z}))=(q_{1}^{*},q_{2}^{*})}, \\
    \Delta D^{*}_{(\ell,m)}(k)&:=\left.\frac{\partial^{(\ell+m)} E[D_{k}|Q_{1}(\mathbf{Z}),Q_{2}(\mathbf{Z})]}{\partial^{\ell} Q_{1}(\mathbf{Z})\partial^{m} Q_{2}(\mathbf{Z})}\right|_{(Q_{1}(\mathbf{Z}),Q_{2}(\mathbf{Z}))=(q_{1}^{*},q_{2}^{*})}.
\end{align*}
\end{theorem}


For any $k\in\{0,1,2\}$ and $\ell,m\in\{0,1,2\}$ such that $\ell+m\leq 2$, we can obtain $\Delta GD^{*}_{(\ell,m)}(k)$ and $\Delta D^{*}_{(\ell,m)}(k)$ by using estimation methods such as local polynomial regression. Note that the density of $\mathbf{V}$ at $(q_{1}^{*},q_{2}^{*})$ is identified as $\Delta D^{*}_{(1,1)}(0)$. For (b) and (c), the second and third terms on the right-hand side correct indirect effects that we discuss in the following.

The identification result of conditional expectations enables us to identify measures of treatment effects.\footnote{In the binary treatment case, by using identification results of conditional expectations, \citet{CL09JE}  propose a semiparametric estimator of the MTE. \citet{BMW17JPE} also employ results to identify MTE with discrete instruments.} For example, if we set $G(Y)=Y$ as we did previously, we obtain
\begin{equation*}
E[Y_{k}-Y_{j}|V_{1}=q_{1}^{*},V_{2}=q_{2}^{*}]  \quad \text{for $k,j\in\{0,1,2\}$ and $k\neq j$.}
\end{equation*}
If we let $G(Y)=\mathbbm{1}(Y\leq y)$ for $y\in\mathbb{R}$, we can identify
\begin{equation*}
    F_{Y_{k}|V_{1}=q_{1}^{*},V_{2}=q_{2}^{*}}(y),\ \ F_{Y_{j}|V_{1}=q_{1}^{*},V_{2}=q_{2}^{*}}(y) \quad \text{for $k,j\in\{0,1,2\}$ and $k\neq j$.}
\end{equation*}
If $F_{Y_{1}|V_{1},V_{2}}$ and $F_{Y_{2}|V_{1},V_{2}}$ are invertible, we identify the quantile treatment effect by taking the difference between the two, that is,
\begin{equation*}
    Q_{Y_{k}|V_{1}=q_{1}^{*},V_{2}=q_{2}^{*}}(\tau)-Q_{Y_{j}|V_{1}=q_{1}^{*},V_{2}=q_{2}^{*}}(\tau) \quad \text{for $k,j\in\{0,1,2\}$ and $k\neq j$,}
\end{equation*}
where $\tau\in(0,1)$.

\begin{figure}[htb]
\hspace{-13mm}
\includegraphics[scale=0.45]{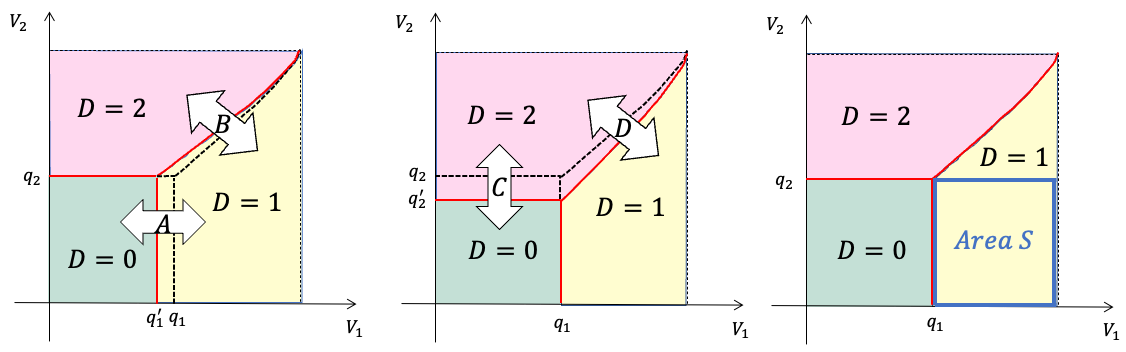}
\caption{This figure depicts flows generated by marginal changes in $Q_{1}(\mathbf{Z})$ and $Q_{2}(\mathbf{Z})$.}
\label{figure_flow}
\end{figure}

Our identification strategy consists of marginal changes in conditional expectations and proper corrections for indirect effects caused by a nonlinear threshold. We can identify $E[G(Y_{0})|V_{1},V_{2}]$ without being troubled by a nonlinear threshold. For instance, when $Q_{2}(\mathbf{Z})$ is fixed, the change in $Q_{1}(\mathbf{Z})$ corresponds exactly to the flow between treatment 0 and 1 over values of $V_{2}$ in $(0,q_{2})$ (flow A in Figure \ref{figure_flow}). Further, taking its derivative at $q_{2}$ provides the flow between treatment 0 and 2 at $(q_{1},q_{2})$ (flow C in Figure \ref{figure_flow}). Therefore, we can identify $E[G(Y_{0})|V_{1},V_{2}]$ by taking derivatives of $E[G(Y)D_{0}|Q_{1}(\mathbf{Z}), Q_{2}(\mathbf{Z})]$ and $E[D_{0}|Q_{1}(\mathbf{Z}), Q_{2}(\mathbf{Z})]$ with respect to $Q_{1}(\mathbf{Z})$ and $Q_{2}(\mathbf{Z})$ as in the binary case.






However, a nonlinear threshold in this model complicates identifications of conditional expectations about treatments 1 and 2.  In order to identify expectations about treatment 1 and treatment 2  conditional on $V_1$ and $ V_2$, we need to take derivatives of $E[G(Y)D_{k}|Q_{1}(\mathbf{Z}), Q_{2}(\mathbf{Z})]$ and $E[D_{k}|Q_{1}(\mathbf{Z}), Q_{2}(\mathbf{Z})]$ for $k=1$ and $2$ with respect to $Q_{1}(\mathbf{Z})$ and $Q_{2}(\mathbf{Z})$  as in the case of treatment 0.  Because marginal changes in $Q_{1}(\mathbf{Z})$ and $Q_{2}(\mathbf{Z})$ only indirectly affect the preference between treatment 1 and 2, we need to deal with indirect changes between treatment 1 and 2.

For instance, when identifying the conditional expectations of treatment 1, we first study area S in Figure \ref{figure_flow}. Because $V_{1}$ is larger than $Q_{1}(\mathbf{Z})$ and $V_{2}$ is smaller than $Q_{2}(\mathbf{Z})$, this area represents those who have the preference order of treatment 1, 0, 2 from top to bottom. When $Q_{1}(\mathbf{Z})$ decreases through $R_{1}(\mathbf{Z})$, people with treatment 0 will move into area S (flow A in Figure \ref{figure_flow}), while this change in $R_1(\mathbf{Z})$ does not affect preferences over treatment 0 and 2. Moreover, when $Q_{2}(\mathbf{Z})$ slightly decreases through $R_{2}(\mathbf{Z})$, people in area S will change their preference orders from treatment 1, 0, 2 to treatment 1, 2, 0. Therefore, confining to area S, we can identify $E[G(Y_{1})|V_{1},V_{2}]$ by taking derivatives of $E[G(Y)D_{1}|Q_{1}(\mathbf{Z}), Q_{2}(\mathbf{Z})]$ and $E[D_{1}|Q_{1}(\mathbf{Z}), Q_{2}(\mathbf{Z})]$  with respect to $Q_1(\mathbf{Z})$ and $Q_2(\mathbf{Z})$ as in treatment 0.




A decrease in $Q_{1}(\mathbf{Z})$ generates another flow, flow B in Figure \ref{figure_flow}. We must remove the marginal changes in treatment 1 caused by flow B. In this case, when $Q_{2}(\mathbf{Z})$ decreases through $R_{2}(\mathbf{Z})$, some individuals taking treatment 1 will move to treatment 2 (flow D in Figure \ref{figure_flow}), but there is no flow from treatment 1 to treatment 0 because the change in $R_{2}(\mathbf{Z})$ does not affect preferences over treatment 1 and 0. We can remove the marginal change between treatment 1 and treatment 0 by substituting flow D for flow B. Then, we can obtain marginal changes generated by $Q_{1}(\mathbf{Z})$ in area S by subtracting flow B from marginal changes generated by $Q_{1}(\mathbf{Z})$ in treatment 1. Therefore, we achieve the identification of $E[G(Y_{1})|V_{1},V_{2}]$.

\section{Identification of thresholds $\mathbf{Q(Z)}$}




In this section, we consider the identification result of $Q_{1}(\mathbf{Z})$ and $Q_{2}(\mathbf{Z})$ that enable us to estimate MTE with results in Theorems \ref{the_iden}. While the propensity score $P(\mathbf{Z})$ plays a role as a threshold in deciding on the treatment in the binary case, $\mathbf{Q(Z)}$ has more complicated relationships with propensity scores $E[D_{0}|\mathbf{Z}]$, $E[D_{1}|\mathbf{Z}]$ and $E[D_{2}|\mathbf{Z}]$. Because each marginal distribution of $V_{1}$ and $V_{2}$ is uniformly distributed on $(0,1)$, $Q_{1}(\mathbf{Z})$ and $Q_{2}(\mathbf{Z})$ correspond to the probability that treatment 0 is preferable to treatment 1 and the probability that treatment 0 is preferable to treatment 2, respectively. Hence, we cannot directly identify $\mathbf{Q}(\mathbf{Z})$ from $E[D_{0}|\mathbf{Z}]$, $E[D_{1}|\mathbf{Z}]$ and $E[D_{2}|\mathbf{Z}]$ because propensity scores only reveal probabilities of each treatment given $\mathbf{Z}$. \footnote{If we have the data of preferences over the choice set as in the case of \citet{KLM16QJE}, we can directly identify $\mathbf {Q}(\mathbf{Z})$.}



 We provide a sufficient condition for the nonparametric identification of $Q_{1}(\mathbf{Z})$ and $Q_{2}(\mathbf{Z})$ that enables us to estimate MTE with results in Theorems \ref{the_iden}. A sufficient condition requires the existence of at least one instrument that significantly and negatively affects only one utility. Let $Z^{[\ell]}$ denote the $\ell$-th component of $\mathbf{Z}$. Let $\mathbf{Z}^{[-\ell]}$ be all the instruments except for the $\ell$-th component. 
\begin{assumption}\label{ass_iden} 
For $k=1,2$, there exists at least one element of $\mathbf{Z}$, say $Z^{[\ell_{k}]}$, and at least one value $a^{[\ell_{k}]}$, such that, given any $\mathbf{z}^{[-\ell_{k}]}\in\mathbf{Z}^{[-\ell_{k}]}$, 
\begin{equation*}
\lim_{z^{[\ell_{k}]}\rightarrow a^{[\ell_{k}]}}R_{k}(z^{[\ell_{k}]},\mathbf{z}^{[-\ell_{k}]})=\infty\ \ \ 
\end{equation*}
and $R_{j}(\mathbf{Z})$ is constant for $j\neq k$.
\end{assumption}
Assumption \ref{ass_iden} imposes a type of exclusion restriction. Conditional on all the regressors except $Z^{[\ell_{k}]}$, one can vary $R_{k}(\mathbf{Z})$ independently. Moreover, we assume the existence of one value $a^{[\ell_{k}]}$ such that, as $z^{[\ell_{k}]}$ approaches $a^{[\ell_{k}]}$, the value of the function $-R_{k}(\mathbf{Z})$ becomes sufficiently small given any $\mathbf{z}^{-[\ell_{k}]}$.

\begin{theorem}\label{the_idenq}
Assume Assumptions \ref{ass_m3} to \ref{ass_V3} and \ref{ass_iden} hold. Then, $Q_{1}(\mathbf{Z}),Q_{2}(\mathbf{Z})$ are identified as
\begin{align*}
    \lim_{z^{[\ell_{2}]}\rightarrow a^{[\ell_{2}]}}H(\mathbf{Z})&=Q_{1}(\mathbf{Z}), \\
    \lim_{z^{[\ell_{1}]}\rightarrow a^{[\ell_{1}]}}H(\mathbf{Z})&=Q_{2}(\mathbf{Z}),
\end{align*}
where
\begin{equation*}
    H(\mathbf{Z}):=\Pr(D_{1}=1|\mathbf{Z})=F_{\mathbf{V}}(Q_{1}(\mathbf{Z}),Q_{2}(\mathbf{Z})).
\end{equation*}
\end{theorem}

The strategy for the identification of $\mathbf{Q}(\mathbf{Z})$ in Theorem \ref{the_idenq} deeply depends on the reduction to the binary treatment setting. As $z^{[\ell_{2}]}$ converges to $a^{[\ell_{2}]}$, for instance, $Q_{2}(\mathbf{Z})$ approximately approaches to 1, which implies that individuals take treatment 0 or treatment 1. Hence, we identify $Q_{1}(\mathbf{Z})$ as in the binary case.

 Assumption \ref{ass_iden} is similar to assumptions for identifying thresholds in the existing literature. \citet{LS18ecta} establish the general identification result of conditional expectations given that $\mathbf{Q}(\mathbf{Z})$ is known. They study the identification of $\mathbf{Q}(\mathbf{Z})$ for some choice models, using the information of these models. Especially, they impose a similar large support assumption for the identification of $\mathbf{Q(Z)}$ in a double hurdle model. Because treatment 0 has the form of a double hurdle model, Theorem \ref{the_idenq} can be considered as the identification result of thresholds in a double hurdle model, as in Theorem 4.2 in \citet{LS18ecta}.

\section{Comparison with the Existing Literature}

 In this section, we briefly review the existing literature regarding the identification of MTE with multivalued treatments and compare them with our results.

\subsection{Lee and Salani\'e (2018)}\label{LS-comaprison}
\citet{LS18ecta} employ the following model:
\begin{equation*}
    D_{k}=d_{k}(\mathbf{V},\mathbf{Q(Z)})
\end{equation*}
where
\begin{align}
d_{k}(\mathbf{V},\mathbf{Q(Z)})&=\sum_{l\in\mathcal{L}}c_{l}^{k}\prod_{j\in l}S_{j}(\mathbf{V},\mathbf{Q(Z)})=\sum_{l\in\mathcal{L}}c_{l}^{k}\prod_{m=1}^{|l|}S_{l_{m}}(\mathbf{V},\mathbf{Q(Z)}), \label{model_LS} 
\end{align}
and $c_{l}^{k}$ is an integer.  Let $\mathcal{J}$ be the set of choices, $\{1,\cdots,J\}$, and let $\mathcal{L}$ be the set of all the subsets of $\mathcal{J}$. Model (\ref{model_LS}) can express any decision model that comprises sums, products, and differences of their indicator functions $S_{j}$.




In Section 5.2,  \citet{LS18ecta}  apply their main theorem to the multiple discrete choice model. Example 5.2 of \citet{LS18ecta} analyzes three treatments, $\mathcal{K}=\{0,1,2\}$. They define
\begin{align}
\tilde{R}_{0,1}(\mathbf{Z})&:={R}_{0}(\mathbf{Z})-{R}_{1}(\mathbf{Z}), \ \ 
&\tilde{R}_{0,2}(\mathbf{Z})&:={R}_{0}(\mathbf{Z})-{R}_{2}(\mathbf{Z}), \ \ 
&\tilde{R}_{1,2}(\mathbf{Z})&:={R}_{1}(\mathbf{Z})-{R}_{2}(\mathbf{Z}), \notag \\
\tilde{U}_{0,1}&:={U}_{0}-{U}_{1}, \ \ &\tilde{U}_{0,2}&:={U}_{0}-{U}_{2}, \ \ 
&\tilde{U}_{1,2}&:={U}_{1}-{U}_{2}. \label{def_RU}
\end{align}
Subsequently, they define
\begin{align}
Q_{0,1}(\mathbf{Z})&:=F_{\tilde{U}_{0,1}}(\tilde{R}_{0,1}(\mathbf{Z})), \ \ 
&Q_{0,2}(\mathbf{Z})&:=F_{\tilde{U}_{0,2}}(\tilde{R}_{0,2}(\mathbf{Z})), \ \ 
&Q_{1,2}(\mathbf{Z})&:=F_{\tilde{U}_{1,2}}(\tilde{R}_{1,2}(\mathbf{Z})), \notag  \\
V_{0,1}&:=F_{\tilde{U}_{0,1}}(\tilde{U}_{0,1}), \ \ 
&V_{0,2}&:=F_{\tilde{U}_{0,2}}(\tilde{U}_{0,2}), \ \ 
&V_{1,2}&:=F_{\tilde{U}_{1,2}}(\tilde{U}_{1,2}). \label{def_QV}
\end{align}
Based on the comparison among utilities as we did in Section \ref{Model and Basic Assumptions}, they define treatments as follows:
 \begin{itemize}
  \item $D=0$ iff $V_{0,1}<Q_{0,1}(\mathbf{Z})$ and                                                                     $V_{0,2}<Q_{0,2}(\mathbf{Z})$,
  \item $D=1$ iff $V_{0,1}>Q_{0,1}(\mathbf{Z})$ and $V_{1,2}<Q_{1,2}(\mathbf{Z})$,
  \item $D=2$ iff $V_{0,2}>Q_{0,2}(\mathbf{Z})$ and $V_{1,2}>Q_{1,2}(\mathbf{Z})$.
 \end{itemize}
Evidently, this corresponds to the decision rule based on model (\ref{model_H}).

\citet{LS18ecta} argue that their main theorems (Theorem 3.1 and Theorem A.1) can identify the MTE if $Q_{0,1}(\mathbf{Z}),Q_{0,2}(\mathbf{Z}),Q_{1,2}(\mathbf{Z})$ are identified and their Assumptions 2.1--2.2 and 3.2--3.4 hold. From this result, they state that we can identify MTE without monotonicity. Moreover, because they identify the MTE via multidimensional cross-derivatives, they do not rely on the identification-at-infinity strategy. 

\begin{figure}[htbp]
\centering
\includegraphics[height=80mm, width=80mm]{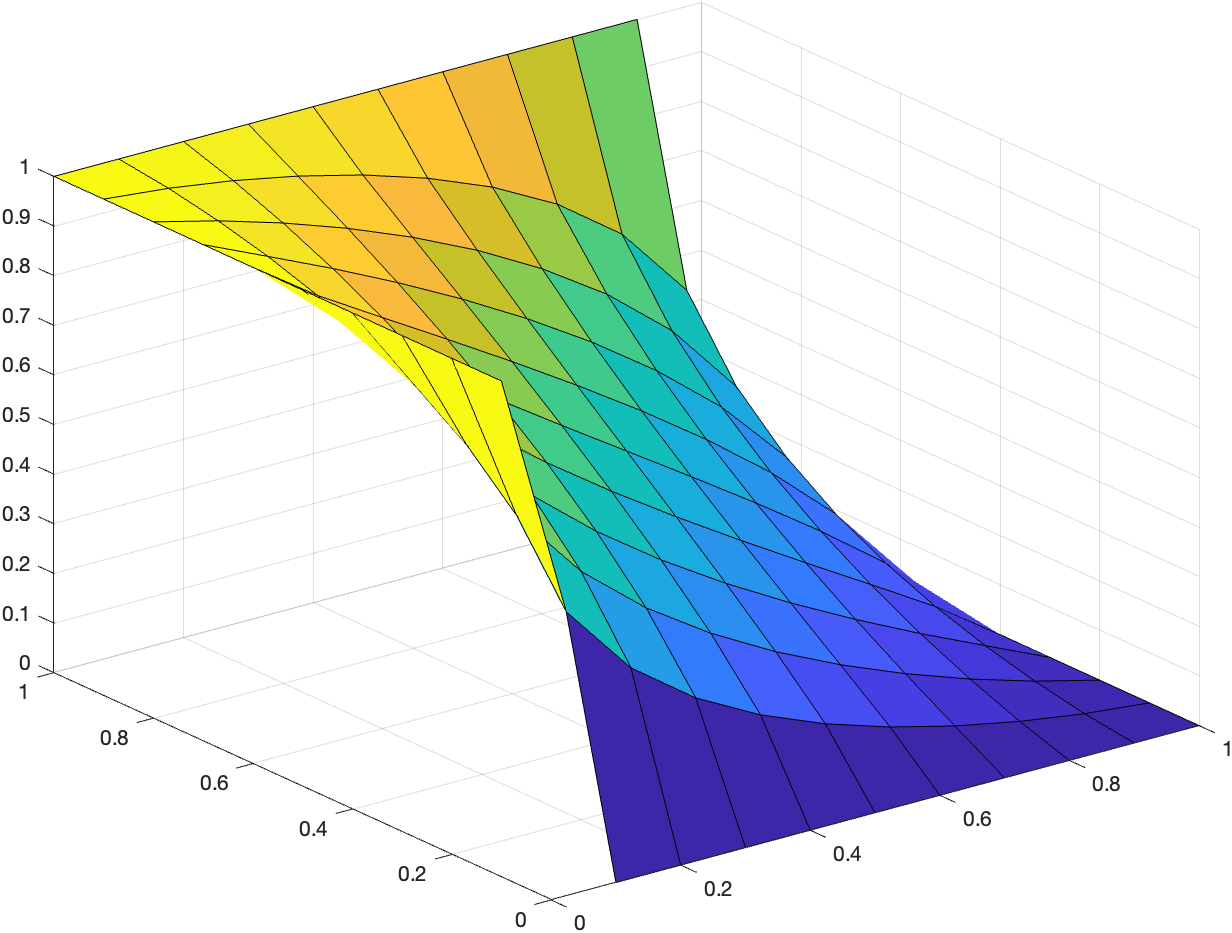}
\caption{This figure shows the support of $\mathbf{V}$ when $U_{0}\sim N(0,0.5)$, $U_{1}\sim N(1,1)$ $U_{2}\sim N(-1,1)$ and $(U_{0}, U_{1}, U_{2})^{'}$ are mutually independent. The right bottom axis, the left bottom axis and the vertical axis correspond to the values of $V_{0,1},V_{0,2}$ and $V_{1,2}$, respectively.}
\label{mathcal_V}
\end{figure}

The following discussion shows that the model in Section 5.2 of \citet{LS18ecta} may not be sufficient to identify the MTE. When we set $\mathbf{V}:=(V_{0,1},V_{0,2}, V_{1,2})^{'}$, by construction we have
\begin{align*}
    F_{\tilde{U}_{0,1}}^{-1}(V_{0,1})&=\tilde{U}_{0,1}, \\ 
&=\tilde{U}_{0,2}-\tilde{U}_{1,2}, \\
&=F_{\tilde{U}_{0,2}}^{-1}(V_{0,2})-F_{\tilde{U}_{1,2}}^{-1}(V_{1,2}).
\end{align*}
 This equality suggests that even if $\mathbf{V}$ is absolutely continuous with respect to the Lebesgue measure on $\mathbb{R}^{3}$, its support is not equal to $[0,1]^{3}$ as in Figure \ref{mathcal_V}. Consequently, $\mathbf{V}$ cannot satisfy Assumption 3.2 in \citet{LS18ecta}, which requires that the joint distribution of $\mathbf{V}$ be absolutely continuous on $\mathbb{R}^{3}$ and that its support be equal to $[0,1]^{3}$.

Our model can be regarded as a double hurdle model for treatment 0. However, in a double hurdle model with three choices, we cannot define some treatments through two thresholds. For example, when treatment 0 has the form of a double hurdle model, the information in $\mathbbm{1}\{V_1 < Q_1(Z)\}$ and $\mathbbm{1}\{V_2 < Q_2(Z)\}$ is not sufficient to determine whether the agent receives either treatment 1 or treatment 2. As a result, we cannot identify the MTE with multivalued treatments. With the introduction of $S_{3}$, we successfully specify $D_{1}$ and $D_{2}$ in our model. Our model cannot be expressed in the form of model (\ref{model_LS}), however, because $S_{3}$ includes the nonlinear transformation of $(V_{1},V_{2})$ and $(Q_{1}(\mathbf{Z}), Q_{2}(\mathbf{Z}))$. Consequently,  Theorem 3.1 in \citet{LS18ecta} is not sufficient to identify the corresponding conditional expectations of $Y_{1}$ or $Y_{2}$ in our model.

\subsection{Heckman and Vytlacil (2007) and Heckman et al. (2008)}
 \citet{HV07HE71} and \citet{HRV08AES} expand the LIV approach to model (\ref{model_H}). They study identification conditions of three treatment effects: the treatment effect of one specific choice versus the next best alternative, the treatment effect of one specific group of choices versus the other group, and the treatment effect of one specified choice versus another choice.
 
 

\citet{HV07HE71} and \citet{HRV08AES} establish sufficient conditions for the identification of the following MTE:
\begin{equation}\label{MTE_H}
    E[Y_{k}-Y_{j}|R_{k}(\mathbf{Z})-R_{j}(\mathbf{Z})=\ell,R_{k}(\mathbf{Z})-U_{k}=R_{j}(\mathbf{Z})-U_{j}]
\end{equation}
for any $\ell\in\mathbb{R}$ and $j,k\in \{0,1,2\}$, such that $j\neq k$. To identify MTE \eqref{MTE_H}, they impose a large support assumption in Theorem 8 \citep{HV07HE71} and Theorem 3 \citep{HRV08AES}. This large support assumption implies that one utility function, $R_{k}(\mathbf{Z})$, can take a sufficiently negative value as in Assumption \ref{ass_iden}. Under the large support assumption, they succeed in reducing the model with three alternatives to a binary case and they achieve the identification of MTE \eqref{MTE_H}.

While we impose Assumption \ref{ass_iden} for the identification of thresholds, our identification strategy does not depend on the large support assumption. As a result, we achieve the identification of the MTE with two-dimensional unobserved heterogeneity $(V_{1},V_{2})$. Moreover, while their MTEs are conditioned on unobserved heterogeneity $U_{k}-U_{j}$, we do not require the information of distributions for heterogeneity.

\subsection{Mountjoy (2022)}

\citet{mountjoy2022community} studies the effect of enrollment in 2-year community colleges on upward mobility, such as years of education and future income. With three valued treatments, he disentangles the treatment effect of 2-year college entry versus the other two treatments into two parts: the treatment effect between 2-year college entry and no college and the treatment effect between 2-year and 4-year entry. Using a novel identification approach, he identifies and estimates those treatment effects with the multivalued treatments.


Let $D_{0},D_{2}$ and $D_{4}$ denote no-college treatment, 2-year college entry treatment, and 4-year college entry treatment. He define $Z_{2}$ and $Z_{4}$ as continuous instrumental variables specific to $D_{2}$ and $D_{4}$, respectively and assume $D_{0},D_{2}$ and $D_{4}$ depend on two instruments, i.e., $D_{0}=D_{0}(z_{2},z_{4})$, $D_{2}=D_{2}(z_{2},z_{4})$ and $D_{4}=D_{4}(z_{2},z_{4})$. The observed outcome $D(z_2,z_4)$ is expressed as $D(z_2,z_4)=\sum_{k=0}^{4} kD_{k}(z_{2},z_{4})$.

Because \citet{mountjoy2022community} is interested in a special case of the marginal treatment effect, that is, the effect of marginal policy changes for 2-year entry on the outcome of  2-year colleges, he defines new MTEs with respect to the marginal change of $z_{2}$. Under regularity conditions, he defines and identifies the following two treatment effects with interesting decomposition.
\begin{align}
    MTE_{2\leftarrow 4}(z_{2},z_{4}):=&\lim_{z_{2}^{\prime}\uparrow z_{2}}E[Y_2-Y_{4}|D(z_2^{\prime},z_4)=2, D(z_2,z_4)=4], \label{MTE_M-1} \\
    MTE_{2\leftarrow 0}(z_{2},z_{4}):=&\lim_{z_{2}^{\prime}\uparrow z_{2}}E[Y_2-Y_{0}|D(z_2^{\prime},z_4)=2, D(z_2,z_4)=0], \label{MTE_M-2} \\
    \frac{\frac{\partial E[Y|z_{2},z_{4}]}{\partial Z_{2}}}{\frac{\partial E[D_{2}|z_{2},z_{4}]}{\partial Z_{2}}}=&\omega(z_{2},z_{4}) MTE_{2\leftarrow 0}(z_{2},z_{4})+(1-\omega(z_{2},z_{4}))MTE_{2\leftarrow 4}(z_{2},z_{4}), \notag 
\end{align}
where he defines
\begin{equation*}
    \omega(z_{2},z_{4}):=\frac{-\frac{\partial E[D_0|z_{2},z_{4}]}{\partial Z_{2}}}{\frac{\partial E[D_2|z_{2},z_{4}]}{\partial Z_{2}}}.
\end{equation*}
MTEs \eqref{MTE_M-1} and \eqref{MTE_M-2} reflect the marginal change between treatments induced by the marginal change in $z_{2}$. 



Our MTEs are based on unobserved heterogeneities $V_{1}$ and $V_{2}$, which refer to the preferences of treatment 1 and treatment 2 over treatment 0, respectively. Therefore, MTE \eqref{MTE_T} corresponds to the marginal change in preferences over the choice set and is entirely different from MTEs  \eqref{MTE_M-1} and \eqref{MTE_M-2}. While \citet{mountjoy2022community} does not require a decision model for the identification of his MTE, we set utilities of 2-year and 4-year entry as like model \eqref{model_H}.  Let $R_{2}(\mathbf{Z})$ and $R_{4}(\mathbf{Z})$ denote observed terms of utilities for a 2-year college and a 4-year college, and $U_2$ and $U_4$ denote unobserved terms of utilities for a 2-year college and a 4-year college, respectively. Then, MTEs in \citet{mountjoy2022community} can be expressed in terms of our MTEs as follows,
\begin{align*}
    &MTE_{2\leftarrow 0}(z_{2},z_{4}) \\
    :=&\frac{\int_{F_{U_4}(R_4(z_{4}))}^{1}E[Y_{2}-Y_{0}|V_{1}=F_{U_2}(R_2(z_{2})), V_{2}=v_{2}]f_{\mathbf{V}}(F_{U_2}(R_2(z_{2})),v_{2})dv_{2}}{\int_{F_{U_4}(R_4(z_{4}))}^{1}f_{\mathbf{V}}(F_{U_2}(R_2(z_{2})),v_{2})dv_{2}}, \\
   &MTE_{2\leftarrow 4}(z_{2},z_{4}) \\ :=&\frac{\int_{F_{U_4}(R_4(z_{4}))}^{1}E[Y_2-Y_4|V_{1}=b(v_{2},z_2,z_4),V_{2}=v_{2}]f_{\mathbf{V}}(b(v_{2},z_2,z_4),v_{2})f_{U_{2}}(F_{U_{2}}^{-1}(b(v_{2},z_2,z_4)))dv_{2}}{\int_{F_{U_4}(R_4(z_{4}))}^{1}f_{\mathbf{V}}(b(v_{2},z_2,z_4),v_{2})f_{U_{2}}(F_{U_{2}}^{-1}(b(v_{2},z_2,z_4)))dv_{2}}
\end{align*}
where we define 
\begin{align*}
    b(v_{2},z_2,z_4):=F_{U_{2}}(F_{U_4}^{-1}(v_2)-R_4(z_{4})+R_{2}(z_2)).
\end{align*}

\section{Generalization}

This section generalizes the framework in Section \ref{Model and Basic Assumptions} and \ref{Model and Identification} to identify the MTE for the discrete choice model with more than two treatments.

\subsection{Model and Assumptions}
 In this subsection, we extend the model constructed in Section
 \ref{Model_description}. We set treatment $k\in\mathcal{K}$ as the baseline and consider the identification of the MTE of treatment $k$ versus treatment $j$ for any $j\neq k$ where $j,k\in\mathcal{K}$ and $K\geq3$. 
 
 Define, for each $i\neq k$ in $\mathcal{K}$, 
\begin{align*}
\tilde{R}_{i,k}(\mathbf{Z})&:={R}_{i}(\mathbf{Z})-{R}_{k}(\mathbf{Z}), \ \
\tilde{U}_{i,k}:={U}_{i}-{U}_{k}.
\end{align*}
Assume $\tilde{U}_{i,k}$  are continuously distributed. Let 
\begin{align*}
Q_{i}(\mathbf{Z})&:=F_{\tilde{U}_{i,k}}(\tilde{R}_{i,k}(\mathbf{Z})), \ \ 
V_{i}:=F_{\tilde{U}_{i,k}}(\tilde{U}_{i,k}),\ \ S_{i}=\mathbbm{1}\{V_{i}<Q_{i}(\mathbf{Z})\}.
\end{align*}
By construction, $Q_{i}(\mathbf{Z})$, $V_{i}$ and $S_{i}$ are defined for each $i$ in $\mathcal{K}$ except $k$. Note that
\begin{align*}
V_{i}<Q_{i}(\mathbf{Z})&\Leftrightarrow F_{\tilde{U}_{i,k}}({U}_{i}-{U}_{k})<F_{\tilde{U}_{i,k}}({R}_{i}(\mathbf{Z})-{R}_{k}(\mathbf{Z})), \\
&\Leftrightarrow {U}_{i}-{R}_{i}(\mathbf{Z})<U_{k}-{R}_{k}(\mathbf{Z}),
\end{align*}
Therefore, we obtain $D_{k}=\prod_{i\in\mathcal{K}\backslash \{k\}}S_{i}$.

Define, for each $i\neq j,k$ in $\mathcal{K}$,
\begin{equation*}
    S_{i}^{*}:=\mathbbm{1}\{V_{i}<F_{\tilde{U}_{i,k}}(F_{\tilde{U}_{j,k}}^{-1}(V_{j})-F_{\tilde{U}_{j,k}}^{-1}(Q_{j}(\mathbf{Z}))+F_{\tilde{U}_{j,i}}^{-1}(Q_{j}(\mathbf{Z}))) \}.
\end{equation*}
By construction, we obtain
\begin{align*}
    &V_{i}<F_{\tilde{U}_{i,k}}(F_{\tilde{U}_{j,k}}^{-1}(V_{j})-F_{\tilde{U}_{j,k}}^{-1}(Q_{j}(\mathbf{Z}))+F_{\tilde{U}_{i,k}}^{-1}(Q_{i}(\mathbf{Z}))) \\
    \Leftrightarrow &F_{\tilde{U}_{i,k}}^{-1}(V_{i})-F_{\tilde{U}_{j,k}}^{-1}(V_{j})<F_{\tilde{U}_{i,k}}^{-1}(Q_{i}(\mathbf{Z}))-F_{\tilde{U}_{j,k}}^{-1}(Q_{j}(\mathbf{Z})) \\
    \Leftrightarrow & U_{i}-U_{j}<R_{i}(\mathbf{Z})-R_{j}(\mathbf{Z}) \\
    \Leftrightarrow & U_{i}-R_{i}(\mathbf{Z})<U_{j}-R_{j}(\mathbf{Z}).
\end{align*}
Hence, we have $D_{j}=\prod_{i\in\mathcal{K}\backslash \{j,k\}}S_{i}^{*}(1-S_{j})$ by definition.

We use the same strategy to identify MTE  of treatment $k$ versus treatment $j$ as in Section \ref{Model and Identification}. Assumptions \ref{ass_m_gen} to \ref{new_assumption_dif_multiple} correspond to Assumptions \ref{ass_m3} to \ref{new_assumption_dif}, respectively.

\begin{assumption}\label{ass_m_gen}
For each $i\neq k$ and $\ell\neq k, j$ in $\mathcal{K}$, $\{V_{i}<Q_{i}(\mathbf{Z})\}$ and $\{V_{\ell}<F_{\tilde{U}_{k,\ell}}(F_{\tilde{U}_{k,j}}^{-1}(V_{j})-F_{\tilde{U}_{k,j}}^{-1}(Q_{j}(\mathbf{Z}))+F_{\tilde{U}_{k,\ell}}^{-1}(Q_{\ell}(\mathbf{Z}))) \}$ are measurable sets.
\end{assumption}

\begin{assumption}[Conditional Independence of Instruments]\label{ass_ind_gen}
$Y_{j}, Y_{k}$ and $\mathbf{V}=(V_{0},\cdots, \\ V_{k-1}, V_{k+1},\cdots,V_{K-1})^{'}$ are jointly independent of $\mathbf{Z}$.
\end{assumption}

\begin{assumption}[Continuously Distributed Unobserved Heterogeneity in the Selection Mechanism]\label{ass_V_gen}
 The joint distribution of $(\tilde{U}_{0,k},\cdots, \tilde{U}_{K-1,k})$ is absolutely continuous with respect to the Lebesgue measure on $\mathbb{R}^{K-1}$. 
\end{assumption}

\begin{assumption}[The Existence of the Moments]\label{ass_mom_gen}
$E[|G(Y_{j})|]<\infty$ and $E[|G(Y_{k})|]<\infty$, where $G$ is a measurable function defined on the support $\mathcal{Y}$ of $Y$, which can be discrete, continuous, or multidimensional.
\end{assumption}


\begin{assumption}\label{new_assumption_dif_multiple}
\
\begin{enumerate}[(1).]
    \item \label{diff_q_multiple}
 For $i\in\{k,j\}$, $E[G(Y)D_{i}|\mathbf{Q}(\mathbf{Z})]$ and $E[D_{i}|\mathbf{Q}(\mathbf{Z})]$ are twice differentiable at $\mathbf{q}^{*}$ where we define $\mathbf{Q}(\mathbf{Z}):=(Q_{0}(\mathbf{Z}),\cdots, Q_{k-1}(\mathbf{Z}), Q_{k+1}(\mathbf{Z}),\cdots,  Q_{K-1}(\mathbf{Z}))^{'}$ and $\mathbf{q}^{*}:=(q_{0}^{*},\cdots, q_{k-1}^{*}, q_{k+1}^{*},\cdots,q_{K-1}^{*})^{'}$.
   \item \label{continuity_v_multiple}  For $i\in\{k,j\}$,  $E[G(Y_{i})|\mathbf{V}]$ and $E[D_{i}|\mathbf{V}]$ are continuous on $(0,1)^{K-1}$.  
   \item \label{bounded_multiple} 
   \begin{enumerate}
   \item \label{bounded_moment_multiple}
   For $i\in\{k,j\}$,  $\sup_{\mathbf{v}\in(0,1)^{(K-1)}}E[|G(Y_{i})||\mathbf{V}=\mathbf{v}]$ is finite.
     \item \label{bounded_conditional_multiple}For any $i\in\mathcal{K}^{\backslash\{k\}}$, density functions satisfy the following:
        \begin{align*}
            &\sup_{(\tilde{u}_{0,k},\cdots, \tilde{u}_{K-1,k})\in\mathbb{R}^{K-1}}\frac{f_{\tilde{U}_{0,k},\cdots, \tilde{U}_{K-1,k}}(\tilde{u}_{0,k},\cdots, \tilde{u}_{K-1,k})}{\prod_{\ell\in\mathcal{K}\backslash\{k, i\}}f_{\tilde{U}_{\ell,k}}(\tilde{u}_{\ell,k})}<\infty.
        \end{align*}
   \end{enumerate}
 \end{enumerate}
\end{assumption}



In studying the general case, we need to introduce the following assumption that innocuously holds in the case $K=3$.
\begin{assumption}\label{new_assumption_density_ratio}
There exists at least one treatment $b^{*}$ in $\mathcal{K}^{\backslash \{j,k\}}$ such that the following conditions hold for any $i\in\mathcal{K}^{\backslash \{j,k\}}$:
\begin{enumerate}[(1).]
    \item\label{new_assumption_density_ratio_differentiation} The density ratio $\left.f_{\tilde{U}_{i,k}}(F_{\tilde{U}_{i,k}}^{-1}(q_{i}))\right/ f_{\tilde{U}_{b^{*},k}}(F_{\tilde{U}_{b^{*},k}}^{-1}(q_{b^{*}}))$ is differentiable at $q_{i}^{*}$ and $q^{*}_{b^{*}}$.
    \item \label{new_assumption_density_ratio_ratios} The following terms are known,
    \begin{align*}
        &\frac{f_{\tilde{U}_{i,k}}(F_{\tilde{U}_{i,k}}^{-1}(q_{i}^{*}))}{f_{\tilde{U}_{b^{*},k}}(F_{\tilde{U}_{b^{*},k}}^{-1}(q_{b^{*}}^{*}))},& &\left.\frac{\partial}{\partial q_{i}}\left(\frac{f_{\tilde{U}_{i,k}}(F_{\tilde{U}_{i,k}}^{-1}(q_{i}))}{f_{\tilde{U}_{b^{*},k}}(F_{\tilde{U}_{b^{*},k}}^{-1}(q_{b^{*}}^{*}))}\right)\right|_{q_{i}=q_{i}^{*}},& \\
        &\left.\frac{\partial}{\partial q_{b}^{*}}\left(\frac{f_{\tilde{U}_{i,k}}(F_{\tilde{U}_{i,k}}^{-1}(q_{i}))}{f_{\tilde{U}_{b^{*},k}}(F_{\tilde{U}_{b^{*},k}}^{-1}(q_{b^{*}}^{*}))}\right)\right|_{q_{b}^{*}=q_{b^{*}}^{*}},& &\left.\frac{\partial^{2}}{\partial  q_{b}^{*}\partial q_{i}}\left(\frac{f_{\tilde{U}_{i,k}}(F_{\tilde{U}_{i,k}}^{-1}(q_{i}))}{f_{\tilde{U}_{b^{*},k}}(F_{\tilde{U}_{b^{*},k}}^{-1}(q_{b^{*}}^{*}))}\right)\right|_{(q_{i},q_{b^{*}})=(q_{i}^{*}, q_{b^{*}}^{*})}.&
    \end{align*}
\end{enumerate}
\end{assumption}
Assumption \ref{new_assumption_density_ratio} is less restrictive than the assumption requiring identification of the density ratio between \(\tilde{U}_{i,k}\) and \(\tilde{U}_{j,k}\), as well as its derivative, for each \(i \in \mathcal{K}^{\setminus\{j,k\}}\). This weaker condition is automatically satisfied when \(K = 3\) because the density ratio is equal to 1.

Focusing on the area $D=j$, as in the case of \(K = 3\), a marginal change in \(Q_j\) induces flows not only into the area where \(D = k\) but also into areas where \(D = \ell\) for any \(\ell \in \mathcal{K}^{\setminus \{j,k\}}\). As outlined in Section 3.3, these indirect effects must be individually offset by substituting marginal changes of \(Q_j\) in areas other than \(D = k\) and \(D = j\). When considering substitution effects from each area, the adjustment cost must be scaled by the density ratio specific to that area. However, since the data cannot distinguish  indirect effects from individual areas, these influences are aggregated into a single sum. 

For cases where \(K \geq 4\), it becomes necessary to identify the density ratios specific to each area. By contrast, when \(K = 3\), there is only one remaining area other than \(D = k\) and \(D = j\), and its density ratio can be directly offset. Therefore, this assumption is unnecessary when \(K = 3\).\footnote{Economic sufficient conditions for identifying these density ratios are provided in Appendix \ref{economics_66}.}

\subsection{Identification Result of MTE}


Conditional on the assumption that $\mathbf{Q}(\mathbf{Z})$ is identified, we can identify conditional expectations $E[G(Y_{k})|\mathbf{V}=\mathbf{q}^{*}]$, $E[G(Y_{j})|\mathbf{V}=\mathbf{q}^{*}]$ and $f_{\mathbf{V}}(\mathbf{q}^{*})$ by partially differentiating conditional expectations $E[G(Y)D_{i}|\mathbf{Q}(\mathbf{Z})=\mathbf{q}]$, $E[D_{i}|\mathbf{Q}(\mathbf{Z})=\mathbf{q}]$ and density ratios, $\left.f_{\tilde{U}_{i,k}}(F_{\tilde{U}_{i,k}}^{-1}(q_{i}))\right/f_{\tilde{U}_{b^{*},k}}(F_{\tilde{U}_{b^{*},k}}^{-1}(q_{b^{*}}))$ for $i\in\mathcal{K}^{\setminus\{k,j\}}$.
\begin{theorem}\label{the_iden_gen}
Let Assumptions \ref{ass_m_gen} to \ref{new_assumption_density_ratio} hold. Then, the conditional expectations of $G(Y_{k}),G(Y_{j})$ are given by
\begin{align*}
&E[G(Y_{k})|\mathbf{V}=\mathbf{q}^{*}] \\
=&\left.\frac{\partial E[G(Y)D_{k}|\mathbf{Q}(\mathbf{Z})]}{\partial \mathbf{Q}(\mathbf{Z})}\right|_{\mathbf{Q(Z)}=\mathbf{q}^{*}}\left/\frac{\partial E[D_{k}|\mathbf{Q}(\mathbf{Z})]}{\partial \mathbf{Q}(\mathbf{Z})}\right|_{\mathbf{Q(Z)}=\mathbf{q}^{*}}, \\
&E[G(Y_{j})|\mathbf{V}=\mathbf{q}^{*}] \notag  \\
=& \frac{\partial^{K-2}}{\partial \mathbf{Q}_{-j}(\mathbf{Z})}\left(\frac{\sum_{i\neq k,j}^{K-1}\left(\frac{f_{\tilde{U}_{i,k}}(F_{\tilde{U}_{i,k}}^{-1}(q_{i}))}{f_{\tilde{U}_{b^{*},k}}(F_{\tilde{U}_{b^{*},k}}^{-1}(q_{b^{*}}))}\times\left.\frac{\partial}{\partial q_{i}}E[G(Y)D_{j}|\mathbf{Q}(\mathbf{Z})=\mathbf{q}]\right|_{(\mathbf{q}_{-j}, q_{j})=(\mathbf{q}_{-j}, q_{j}^{*})}  \right)}{\sum_{i\neq k,j}^{K-1}\left(\frac{f_{\tilde{U}_{i,k}}(F_{\tilde{U}_{i,k}}^{-1}(q_{i}))}{f_{\tilde{U}_{b^{*},k}}(F_{\tilde{U}_{b^{*},k}}^{-1}(q_{b^{*}}))}\times\left.\frac{\partial}{\partial q_{i}}E[D_{j}|\mathbf{Q}(\mathbf{Z})=\mathbf{q}]\right|_{(\mathbf{q}_{-j}, q_{j})=(\mathbf{q}_{-j}, q_{j}^{*})}  \right) } \right.  \\
    \times& \left.\left.\frac{\partial}{\partial q_{j}}E[(D_{j}+D_{k})|\mathbf{Q}(\mathbf{Z})=\mathbf{q}]\right|_{(\mathbf{q}_{-j}, q_{j})=(\mathbf{q}_{-j}, q_{j}^{*})} \right. \\
    -&\left.\left.\left.\frac{\partial}{\partial q_{j}}E[G(Y)D_{j}|\mathbf{Q}(\mathbf{Z})=\mathbf{q}]\right|_{(\mathbf{q}_{-j}, q_{j})=(\mathbf{q}_{-j}, q_{j}^{*})}\right)\right|_{\mathbf{q}=\mathbf{q}^{*}} \left/\frac{\partial E[D_{k}|\mathbf{Q}(\mathbf{Z})]}{\partial \mathbf{Q}(\mathbf{Z})}\right|_{\mathbf{Q(Z)}=\mathbf{q}^{*}}
\end{align*}
where we define  $\mathbf{a}_{-i}$ as the vector that removes $a_{i}$ from the original vector, namely
$\mathbf{a}_{-i}=(a_{1},\cdots, a_{i-1}, a_{i+1}, \cdots, a_{n})$. 
\end{theorem}

\section{Conclusion}

We study the identification of MTE with multivalued treatments. Our model is based on a multinomial choice model with utility maximization. We establish sufficient conditions for the identification of marginal treatment effects with multidimensional unobserved heterogeneity, which reveals treatment effects conditioned on the willingness to participate in treatments against a specific treatment. Our MTE generalizes the MTE defined in \citet{HV05ecta} in binary treatment models and our identification strategy does not depend on the large support assumption required by \citet{HV07HE71} and \citet{HRV08AES}.  We also establish a sufficient condition for identifying thresholds. 


\singlespace{ 
\bibliography{main}

\begin{thebibliography}{19}
\newcommand{\enquote}[1]{``#1''}
\expandafter\ifx\csname natexlab\endcsname\relax\def\natexlab#1{#1}\fi

\bibitem[{Angrist and Imbens(1995)}]{AI95JASA}
Angrist, J.~D. and Imbens, G.~W. (1995), \enquote{Two-stage least squares
  estimation of average causal effects in models with variable treatment
  intensity,} \textit{Journal of the American statistical Association}, 90,
  431--442.

\bibitem[{Brinch et~al.(2017)Brinch, Mogstad, and Wiswall}]{BMW17JPE}
Brinch, C., Mogstad, M., and Wiswall, M. (2017), \enquote{Beyond LATE with a
  discrete instrument,} \textit{Journal of Political Economy}, 125, 985--1039.

\bibitem[{Carneiro and Lee(2009)}]{CL09JE}
Carneiro, P. and Lee, S. (2009), \enquote{Estimating distributions of potential
  outcomes using local instrumental variables with an application to changes in
  college enrollment and wage inequality,} \textit{Journal of Econometrics},
  149, 191--208.

\bibitem[{Dahl(2002)}]{D02ECTA}
Dahl, G.~B. (2002), \enquote{Mobility and the return to education: Testing a
  Roy model with multiple markets,} \textit{Econometrica}, 70, 2367--2420.

\bibitem[{Fusejima(2024)}]{F20ar}
Fusejima, K. (2024), \enquote{Identification of multi-valued treatment effects
  with unobserved heterogeneity,} \textit{Journal of Econometrics}, 238,
  105563.

\bibitem[{Heckman and Pinto(2018)}]{HP18ecta}
Heckman, J.~J. and Pinto, R. (2018), \enquote{Unordered Monotonicity,}
  \textit{Econometrica}, 86, 1--35.

\bibitem[{Heckman et~al.(2006)Heckman, Urzua, and Vytlacil}]{HRV06RES}
Heckman, J.~J., Urzua, S., and Vytlacil, E. (2006), \enquote{{Understanding
  Instrumental Variables in Models with Essential Heterogeneity},} \textit{The
  Review of Economics and Statistics}, 88, 389--432.

\bibitem[{Heckman et~al.(2008)Heckman, Urzua, and Vytlacil}]{HRV08AES}
--- (2008), \enquote{Instrumental Variables in Models with Multiple Outcomes:
  the General Unordered Case,} \textit{Annales d'{\'E}conomie et de
  Statistique}, 151--174.

\bibitem[{Heckman and Vytlacil(2005)}]{HV05ecta}
Heckman, J.~J. and Vytlacil, E. (2005), \enquote{Structural Equations,
  Treatment Effects, and Econometric Policy Evaluation,} \textit{Econometrica},
  73, 669--738.

\bibitem[{Heckman and Vytlacil(1999)}]{HV99NAS}
Heckman, J.~J. and Vytlacil, E.~J. (1999), \enquote{Local Instrumental
  Variables and Latent Variable Models for Identifying and Bounding Treatment
  Effects,} \textit{Proceedings of the National Academy of Sciences of the
  United States of America}, 96, 4730--4734.

\bibitem[{Heckman and Vytlacil(2007)}]{HV07HE71}
--- (2007), \enquote{Chapter 71 Econometric Evaluation of Social Programs, Part
  II: Using the Marginal Treatment Effect to Organize Alternative Econometric
  Estimators to Evaluate Social Programs, and to Forecast their Effects in New
  Environments,} in \textit{Handbook of Econometrics}, eds. Heckman, J.~J. and
  Leamer, E.~E., Elsevier, vol.~6, pp. 4875--5143.

\bibitem[{Imbens and Angrist(1994)}]{IA94ecta}
Imbens, G.~W. and Angrist, J.~D. (1994), \enquote{Identification and Estimation
  of Local Average Treatment Effects,} \textit{Econometrica}, 62, 467--475.

\bibitem[{Kirkeboen et~al.(2016)Kirkeboen, Leuven, and Mogstad}]{KLM16QJE}
Kirkeboen, L.~J., Leuven, E., and Mogstad, M. (2016), \enquote{Field of study,
  earnings, and self-selection,} \textit{The Quarterly Journal of Economics},
  131, 1057--1111.

\bibitem[{Kline and Walters(2016)}]{KW16QJE}
Kline, P. and Walters, C.~R. (2016), \enquote{Evaluating public programs with
  close substitutes: The case of Head Start,} \textit{The Quarterly Journal of
  Economics}, 131, 1795--1848.

\bibitem[{Lee and Salani{\'e}(2018)}]{LS18ecta}
Lee, S. and Salani{\'e}, B. (2018), \enquote{Identifying Effects of Multivalued
  Treatments,} \textit{Econometrica}, 86, 1939--1963.

\bibitem[{Matzkin(1993)}]{M93JE}
Matzkin, R. (1993), \enquote{Nonparametric identification and estimation of
  polychotomous choice models,} \textit{Journal of Econometrics}, 58, 137--168.

\bibitem[{McFadden(1974)}]{M74B}
McFadden, D. (1974), \enquote{Conditional logit analysis of qualitative choice
  behavior,} \textit{Frontiers in Econometrics}, 105--142.

\bibitem[{Mountjoy(2022)}]{mountjoy2022community}
Mountjoy, J. (2022), \enquote{Community colleges and upward mobility,}
  \textit{American Economic Review}, 112, 2580--2630.

\bibitem[{Vytlacil(2002)}]{V02ecta}
Vytlacil, E. (2002), \enquote{Independence, Monotonicity, and Latent Index
  Models: An Equivalence Result,} \textit{Econometrica}, 70, 331--341.

\end{thebibliography}
}


\indent


\appendix

\renewcommand{\theequation}{A.\arabic{equation}} 
\def\thesection{\Alph{section}}
\setcounter{equation}{0}

{\LARGE{\textbf{Appendix}}}

\section{Proofs and Auxiliary Results}

\subsection{Proof of Theorem \ref{the_iden}}



\begin{proof}[Proof of Theorem \ref{the_iden}]
We first show part (a), namely, 
\begin{align}
&E[G(Y_{0})|V_{1}=q_{1}^{*},V_{2}=q_{2}^{*}] \notag \\
=&\left.\frac{\partial^{2} E[G(Y)D_{0}|Q_{1}(\mathbf{Z}),Q_{2}(\mathbf{Z})]}{\partial Q_{1}(\mathbf{Z})\partial Q_{2}(\mathbf{Z})}\right|_{(Q_{1}(\mathbf{Z}),Q_{2}(\mathbf{Z}))=(q_{1}^{*},q_{2}^{*})}\left/ \frac{\partial^{2} E[D_{0}|Q_{1}(\mathbf{Z}),Q_{2}(\mathbf{Z})]}{\partial Q_{1}(\mathbf{Z})\partial Q_{2}(\mathbf{Z})}\right|_{(Q_{1}(\mathbf{Z}),Q_{2}(\mathbf{Z}))=(q_{1}^{*},q_{2}^{*})}. \label{target_1-1} 
\end{align}  
 Let $\mathbf{Q}(\mathbf{Z})$ and $\mathbf{V}$ denote a vector of thresholds $(Q_{1}(\mathbf{Z}),Q_{2}(\mathbf{Z}))^{\prime}$  and heterogeneity $(V_{1},V_{2})^{\prime}$. Define $d_{0}(\mathbf{V},\mathbf{Q(Z)})$ as the indicator for treatment 0, i.e. $d_{0}(\mathbf{V},\mathbf{Q(Z)}):=\mathbbm{1}\{V_{1}<Q_{1}(\mathbf{Z})\}\times\mathbbm{1}\{V_{2}<Q_{2}(\mathbf{Z})\}$. Set $\mathbf{q}=(q_{1},q_{2})^{\prime}$. Under assumptions of the theorem, for any $\mathbf{q}$ in the range of $\mathbf{Q(Z)}$, we obtain
\begin{align}
&E[G(Y)D_{0}|\mathbf{Q}(\mathbf{Z})=\mathbf{q}] \notag \\
=&E[G(Y_{0})|D_{0}=1,\mathbf{Q}(\mathbf{Z})=\mathbf{q}]\Pr[D_{0}=1|\mathbf{Q}(\mathbf{Z})=\mathbf{q}] \notag \\
=&E[G(Y_{0})|d_{0}(\mathbf{V},\mathbf{Q(\mathbf{Z})})=1,\mathbf{Q(Z)=q}]\Pr[d_{0}(\mathbf{V},\mathbf{Q(\mathbf{Z})})=1|\mathbf{Q(Z)=q}] \notag \\
=&E[G(Y_{0})|d_{0}(\mathbf{V},\mathbf{q})=1]\Pr[d_{0}(\mathbf{V},\mathbf{q})=1] \notag \\
=&E[G(Y_{0})\mathbbm{1}(d_{0}(\mathbf{V},\mathbf{q})=1)] \notag \\
=&E[E[G(Y_{0})|\mathbf{V}]\mathbbm{1}(d_{0}(\mathbf{V},\mathbf{q})=1)], \label{d1_equ}
\end{align} 
where the third equality holds by Assumption \ref{ass_ind3}. Hence, for any $(q_{1},q_{2})^{\prime}$ in the range of $\mathbf{Q}(\mathbf{Z})$, we have
\begin{equation}\label{D0_fubini}
E[G(Y)D_{0}|\mathbf{Q({Z})}=\mathbf{q}]=\int E[G(Y_{0})|\mathbf{V}=\mathbf{v}]g(\mathbf{v};\mathbf{q})f_{\mathbf{V}}(\mathbf{v})d\mathbf{v}   
\end{equation}
where $\mathbf{v}=(v_{1},v_{2})^{\prime}$ and $g(\mathbf{v};\mathbf{q}):=\mathbbm{1}\{v_{1}<q_{1}\}\mathbbm{1}\{v_{2}<q_{2}\}$. From Fubini's theorem, the right-hand side (RHS) of \eqref{D0_fubini} is written as 
\begin{equation*}
\int_{0}^{q_{2}}\int_{0}^{q_{1}}E[G(Y_{0})|V_{1}=v_{1},V_{2}=v_{2}]f_{\mathbf{V}}(v_{1},v_{2})dv_{1}dv_{2}.
\end{equation*}



It follows from Assumption \ref{new_assumption_dif} \eqref{continuity_v} and the Leibniz integral rule that, for any $v_{2}\in(0,1)$
\begin{equation*}
    \left.\frac{\partial}{\partial q_{1}}\int_{0}^{q_{1}}E[G(Y_{0})|V_{1}=v_{1},V_{2}=v_{2}]f_{\mathbf{V}}(v_{1},v_{2})dv_{1}\right|_{q_{1}=q_{1}^{*}}=E[G(Y_{0})|V_{1}=q_{1}^{*}, V_{2}=v_{2}]f_{\mathbf{V}}(q_{1}^{*},v_{2}).
\end{equation*}
Furthermore, because $\sup_{(v_{1},v_{2})\in(0,1)^{2}}E[|G(Y_{0})||V_{1}=v_{1}, V_{2}=v_{2}]$ is finite from Assumption \ref{new_assumption_dif} \eqref{bounded_moment}, 
and $f_{\mathbf{V}}(q_{1}^{*},v_{2})$ is integrable with respect to $v_{2}$, we can exchange differentiation and integral to obtain
\begin{equation}\label{D0_outcome_q_1_derivative}
    \left.\frac{\partial}{\partial q_{1}}E[G(Y)D_{0}|Q_{1}(\mathbf{Z})=q_{1},Q_{2}(\mathbf{Z})=q_{2}]\right|_{q_{1}=q_{1}^{*}}=\int_{0}^{q_{2}}E[G(Y_{0})|V_{1}=q_{1}^{*}, V_{2}=v_{2}]f_{\mathbf{V}}(q_{1}^{*},v_{2})dv_{2}.
\end{equation}
By applying the Leibniz integral rule to the above equation with respect to $q_{2}$, we have
\begin{equation}\label{D0_outcome_result}
\left.\frac{\partial^{2}}{\partial q_{1} \partial q_{2}}E[G(Y)D_{0}|\mathbf{Q}(\mathbf{Z})=\mathbf{q}]\right|_{(q_{1},q_{2})=(q_{1}^{*},q_{2}^{*})}=E[G(Y_{0})|V_{1}=q_{1}^{*},V_{2}=q_{2}^{*}]f_{\mathbf{V}}(q_{1}^{*},q_{2}^{*}).
\end{equation}


We proceed to show the identification result of the density of $\mathbf{V}$, namely, 
\begin{equation}
         f_{\mathbf{V}}(q_{1}^{*},q_{2}^{*})=\left.\frac{\partial^{2} E[D_{0}|Q_{1}(\mathbf{Z}),Q_{2}(\mathbf{Z})]}{\partial Q_{1}(\mathbf{Z})\partial Q_{2}(\mathbf{Z})}\right|_{(Q_{1}(\mathbf{Z}),Q_{2}(\mathbf{Z}))=(q_{1}^{*},q_{2}^{*})}. \label{target_1-2}
\end{equation}
From a similar argument to ($\ref{d1_equ}$), under assumptions of the
theorem, for any $(q_{1},q_{2})^{\prime}$ in the range of $\mathbf{Q}(\mathbf{Z})$, we obtain
\begin{equation*}
E[D_{0}|\mathbf{Q}(\mathbf{Z})=\mathbf{q}]=E[\mathbbm{1}(d_{0}(\mathbf{V},\mathbf{q})=1)]
=\int g(\mathbf{v};\mathbf{q})f_{\mathbf{V}}(\mathbf{v})d\mathbf{v}.
\end{equation*}
Therefore, from the same argument as the one leading to \eqref{D0_outcome_result},  \eqref{target_1-2} holds. The required result (\ref{target_1-1}) follows from  \eqref{D0_outcome_result} and   \eqref{target_1-2}.




 We move on to the proof of part (b). Let $d_{1}(\mathbf{V},\mathbf{Q(\mathbf{Z})})$ denote the indicator for treatment 1, i.e.  $d_{1}(\mathbf{V},\mathbf{Q(\mathbf{Z})})=\mathbbm{1}\{F_{U_{2}}(F_{U_{1}}^{-1}(V_{1})-F_{U_{1}}^{-1}(Q_{1}(\mathbf{Z}))+F_{U_{2}}^{-1}(Q_{2}(\mathbf{Z})))\geq V_{2}\}\times\mathbbm{1}\{V_{1}\geq Q_{1}(\mathbf{Z})\}$.  From a similar argument to ($\ref{d1_equ}$), under assumptions of the theorem, for any $(q_{1},q_{2})^{\prime}$ in the range of $\mathbf{Q}(\mathbf{Z})$,  we obtain
\begin{align}
E[G(Y)D_{1}|\mathbf{Q}(\mathbf{Z})=\mathbf{q}]=&E[E[G(Y_{1})|\mathbf{V}]\mathbbm{1}(d_{1}(\mathbf{V},\mathbf{q})=1)] \notag \\
=&\int E[G(Y_{1})|\mathbf{V}=\mathbf{v}]\ell(\mathbf{v};\mathbf{q})f_{\mathbf{V}}(\mathbf{v})d\mathbf{v}, \label{D1_fubini}
\end{align}
where $\ell(\mathbf{v};\mathbf{q})=\mathbbm{1}\left\{v_{1}\geq q_{1}\right\}\mathbbm{1}\{F_{U_{2}}(F_{U_{1}}^{-1}(v_{1})-F_{U_{1}}^{-1}(q_{1})+F_{U_{2}}^{-1}(q_{2}))\geq v_{2}\}$. From Fubini's theorem, the RHS of \eqref{D1_fubini} is written as 
\begin{equation}
\int_{q_{1}}^{1}\int_{0}^{F_{U_{2}}(F_{U_{1}}^{-1}(v_{1})-F_{U_{1}}^{-1}(q_{1})+F_{U_{2}}^{-1}(q_{2}))}E[G(Y_{1})|V_{1}=v_{1},V_{2}=v_{2}]f_{\mathbf{V}}(v_{1},v_{2})dv_{2}dv_{1}. \label{D_1_Fubini}
\end{equation}


For treatment a, b, and c, let $a\succ b\succ c$ denote the preference order, treatment a, b, and c. Let $\mathcal{V}:=(0,1)^{2}$ denote the support of $\mathbf{V}$, and let $\mathcal{V}\{a\succ b\succ c\}$ denote the support of $\mathbf{V}$ with the preference order $a\succ b\succ c$. In the following, let $q_{1}$ and $q_{2}$ denote arbitrary points of neighborhoods of $q_{1}^{*}$ and $q_{2}^{*}$, respectively.

We decompose the domain of integration in \eqref{D_1_Fubini} into two areas, $\mathcal{V}\{1\succ 2 \succ 0\}$, and $\mathcal{V}\{1\succ 0 \succ 2\}$,
\begin{align}
   &E[G(Y)D_{1}|\mathbf{Q}(\mathbf{Z})=\mathbf{q}]\notag  \\=&\int_{q_{1}}^{1}\int_{0}^{a(v_{1},q_{1},q_{2})}E[G(Y_{1})|V_{1}=v_{1},V_{2}=v_{2}]f_{\mathbf{V}}(v_{1},v_{2})dv_{2}dv_{1} \label{D_1_original_step1} \\
    =&A(q_{1},q_{2})+B(q_{1},q_{2}) \notag
\end{align}
where
\begin{align*}
A(q_{1},q_{2})&:=\int_{0}^{q_{2}}\int_{q_{1}}^{1}E[G(Y_{1})|V_{1}=v_{1},V_{2}=v_{2}]f_{\mathbf{V}}(v_{1},v_{2})dv_{1}dv_{2}, \\
B(q_{1},q_{2})&:=\int_{q_{2}}^{1}\int_{b(v_{2},q_{1},q_{2})}^{1}E[G(Y_{1})|V_{1}=v_{1},V_{2}=v_{2}]f_{\mathbf{V}}(v_{1},v_{2})dv_{1}dv_{2}, \\
a(v_{1},q_{1},q_{2})&:=F_{U_{2}}(F_{U_{1}}^{-1}(v_{1})-F_{U_{1}}^{-1}(q_{1})+F_{U_{2}}^{-1}(q_{2})), \\
    b(v_{2},q_{1},q_{2})&:=F_{U_{1}}(F^{-1}_{U_{2}}(v_{2})-F^{-1}_{U_{2}}(q_{2})+F^{-1}_{U_{1}}(q_{1})).
\end{align*}
$ A(q_{1},q_{2})$ and $B(q_{1},q_{2})$ correspond to $\mathcal{V}\{1\succ0\succ2\}$ and $\mathcal{V}\{1\succ2\succ0\}$, respectively. 

First, we examine the effect of a marginal change in $Q_{1}(\mathbf{Z})$ on $A(q_{1},q_{2})$. Because we can exchange the order of differentiation and integration as in the case of treatment 0, we obtain 
\begin{align}
    \left.\frac{\partial}{\partial q_{1}}A(q_{1},q_{2})\right|_{q_{1}=q_{1}^{*}}=-\int_{0}^{q_{2}}E[G(Y_{1})|V_{1}=q_{1}^{*},V_{2}=v_{2}]f_{\mathbf{V}}(q_{1}^{*},v_{2})dv_{2} . \label{diff_order_1-0-2}
\end{align}



Second, we examine the effect of a marginal change in $Q_{1}(\mathbf{Z})$ on $B(q_{1},q_{2})$. It follows from Assumption \ref{new_assumption_dif} \eqref{diff_q}, \eqref{continuity_v} and  the Leibniz integral rule that, for any $v_{2}\in(0,1)$,
\begin{align*}
    &\left.\frac{\partial}{\partial q_{1}}\int_{b(v_{2},q_{1},q_{2})}^{1}E[G(Y_{1})|V_{1}=v_{1},V_{2}=v_{2}]f_{\mathbf{V}}(v_{1},v_{2})dv_{1}\right|_{q_{1}=q_{1}^{*}} \\
    =&-E[G(Y_{1})|V_{1}=b(v_{2},q_{1}^{*},q_{2}),V_{2}=v_{2}]f_{\mathbf{V}}(b(v_{2},q_{1}^{*},q_{2}),v_{2})\frac{f_{U_{1}}(F_{U_{1}}^{-1}(b(v_{2},q_{1}^{*},q_{2})))}{f_{U_{1}}(F_{U_{1}}^{-1}(q_{1}^{*}))}.
\end{align*}
Through the argument of the change of variables, we have
\begin{equation*}
    f_{\mathbf{V}}(v_{1},v_{2})=\frac{f_{U_{1},U_{2}}(F_{U_{1}}^{-1}(v_{1}),F_{U_{2}}^{-1}(v_{2}))}{f_{U_{1}}(F_{U_{1}}^{-1}(v_{1}))f_{U_{2}}(F_{U_{2}}^{-1}(v_{2}))}.
\end{equation*}
Because $\sup_{(v_{1},v_{2})\in(0,1)^{2}}E[|G(Y_{1})||V_{1}=v_{1}, V_{2}=v_{2}]$ and $\sup_{(u_{1},u_{2})\in\mathbb{R}^{2}}(f_{U_{1},U_{2}}(u_{1},u_{2})/f_{U_{1}}(u_{1}))$ are finite from Assumption \ref{new_assumption_dif} \eqref{bounded}, we can exchange the order of differentiation and integration. Therefore, we obtain 
\begin{align}
    &\left.\frac{\partial}{\partial q_{1}}B(q_{1},q_{2})\right|_{q_{1}=q_{1}^{*}}  \notag \\
=-&\int_{q_{2}}^{1}E[G(Y_{1})|V_{1}=b(v_{2},q_{1}^{*},q_{2}),V_{2}=v_{2}]f_{\mathbf{V}}(b(v_{2},q_{1}^{*},q_{2}),v_{2})\frac{f_{U_{1}}(F_{U_{1}}^{-1}(b(v_{2},q_{1}^{*},q_{2})))}{f_{U_{1}}(F_{U_{1}}^{-1}(q_{1}^{*}))}dv_{2}.  \label{diff_order_1-2-0}
\end{align}

We derive an alternate expansion of the RHS of \eqref{diff_order_1-2-0}. First, we use the change of variables. When we set $v_{2}=a(v_{1},q_{1}^{*},q_{2})$, by definition, we obtain $v_{1}=b(v_{2},q_{1}^{*},q_{2})$.
It holds from \eqref{diff_order_1-2-0} that 
\begin{align}
&\int_{q_{2}
    }^{1}  E[G(Y_{1})|V_{1}=b(v_{2},q_{1}^{*},q_{2}), V_{2}=v_{2}]f_{\mathbf{V}}(b(v_{2},q_{1}^{*},q_{2}), v_{2})f_{U_{1}}(F_{U_{1}}^{-1}(b(v_{2},q_{1}^{*},q_{2})))dv_{2} \notag \\
 =&\int_{q_{2}
    }^{1}  E[G(Y_{1})|V_{1}=b(v_{2},q_{1}^{*},q_{2}), V_{2}=v_{2}]f_{\mathbf{V}}(b(v_{2},q_{1}^{*},q_{2}), v_{2})f_{U_{2}}(F_{U_{2}}^{-1}(v_{2})) \notag \\
    &\times f_{U_{1}}(F_{U_{1}}^{-1}(b(v_{2},q_{1}^{*},q_{2})))\frac{1}{f_{U_{2}}(F_{U_{2}}^{-1}(v_{2}))}dv_{2}  \notag \\
  =&\int_{q_{2}
    }^{1}  E[G(Y_{1})|V_{1}=b(v_{2},q_{1}^{*},q_{2}), V_{2}=v_{2}]f_{\mathbf{V}}(b(v_{2},q_{1}^{*},q_{2}), v_{2})f_{U_{2}}(F_{U_{2}}^{-1}(v_{2})) \notag \\
    &\times f_{U_{1}}(F^{-1}_{U_{2}}(v_{2})-F^{-1}_{U_{2}}(q_{2})+F^{-1}_{U_{1}}(q_{1}^{*}))\frac{dF_{U_{2}}^{-1}(v_{2})}{dv_{2}}
    dv_{2} \notag \\
    =&\int_{q_{2}
    }^{1}  E[G(Y_{1})|V_{1}=b(v_{2},q_{1}^{*},q_{2}), V_{2}=v_{2}]f_{\mathbf{V}}(b(v_{2},q_{1}^{*},q_{2}), v_{2})f_{U_{2}}(F_{U_{2}}^{-1}(v_{2}))\frac{dv_{1}}{dv_{2}}
    dv_{2} \notag \\ =&\int_{q_{1}^{*}}^{1}E[G(Y_{1})|V_{1}=v_{1},V_{2}=a(v_{1},q_{1}^{*},q_{2})]f_{\mathbf{V}}(v_{1},a(v_{1},q_{1}^{*},q_{2}))f_{U_{2}}(F_{U_{2}}^{-1}(a(v_{1},q_{1}^{*},q_{2})))dv_{1}.\label{D1_Step3-3}
\end{align}
Second, we express \eqref{D1_Step3-3} as a function of  a partial derivative of $E[G(Y)D_{1}|\mathbf{Q}(\mathbf{Z})=\mathbf{q}]$ with respect to $Q_{2}(\mathbf{Z})$. It follows from Assumption \ref{new_assumption_dif} (2) and the Leibniz integral rule that, for any $v_{1}\in(0,1)$,
\begin{align*}
    &\left.\frac{\partial}{\partial q_{2}}\int_{0}^{a(v_{1},q_{1},q_{2})}E[G(Y_{1})|V_{1}=v_{1},V_{2}=v_{2}]f_{\mathbf{V}}(v_{1},v_{2})dv_{2}\right|_{q_{2}=q_{2}} \notag \\
    =&E[G(Y_{1})|V_{1}=v_{1},V_{2}=a(v_{1},q_{1},q_{2})]f_{\mathbf{V}}(v_{1},a(v_{1},q_{1},q_{2}))\frac{f_{U_{2}}(F_{U_{2}}^{-1}(a(v_{1},q_{1},q_{2})))}{f_{U_{2}}(F_{U_{2}}^{-1}(q_{2})))}. 
\end{align*}
As in the case of \eqref{diff_order_1-2-0}, we can exchange the order of differentiation and integration. Hence, differentiating the second line of \eqref{D_1_original_step1} with respect to $q_{2}$ gives
\begin{align}
    &\left.\frac{\partial}{\partial q_{2}}E[G(Y)D_{1}|Q_{1}(\mathbf{Z})=q_{1}^{*},Q_{2}(\mathbf{Z})=q_{2}]\right|_{q_{2}=q_{2}} \notag \\
    =&\int_{q_{1}^{*}}^{1}E[G(Y_{1})|V_{1}=v_{1},V_{2}=a(v_{1},q_{1}^{*},q_{2})]f_{\mathbf{V}}(v_{1},a(v_{1},q_{1}^{*},q_{2}))\frac{f_{U_{2}}(F_{U_{2}}^{-1}(a(v_{1},q_{1}^{*},q_{2})))}{f_{U_{2}}(F_{U_{2}}^{-1}(q_{2})))}dv_{1}.  \label{D1_Step2}
\end{align}
In conjunction with \eqref{diff_order_1-2-0}, we obtain
\begin{equation}\label{B_q_1_new}
    \left.\frac{\partial}{\partial q_{1}}B(q_{1},q_{2})\right|_{q_{1}=q_{1}^{*}}= -\left.\frac{\partial}{\partial q_{2}}E[G(Y)D_{1}|Q_{1}(\mathbf{Z})=q_{1}^{*},Q_{2}(\mathbf{Z})=q_{2}]\right|_{q_{2}=q_{2}}\times \frac{f_{U_{2}}(F_{U_{2}}^{-1}(q_{2})))}{f_{U_{1}}(F_{U_{1}}^{-1}(q_{1}^{*}))}.
\end{equation}

We show the likelihood ratio is identifiable through the following equality,
\begin{align}
&\frac{f_{U_{2}}(F_{U_{2}}^{-1}(q_{2})))}{f_{U_{1}}(F_{U_{1}}^{-1}(q_{1}^{*}))}=-\frac{\left.\frac{\partial}{\partial q_{1}}E[(D_{0}+D_{1})|Q_{1}(\mathbf{Z})=q_{1},Q_{2}(\mathbf{Z})=q_{2}]\right|_{q_{1}=q_{1}^{*}}}{\left.\frac{\partial}{\partial q_{2}}E[D_{1}|Q_{1}(\mathbf{Z})=q_{1}^{*},Q_{2}(\mathbf{Z})=q_{2}]\right|_{q_{2}=q_{2}}}.\label{identity-3}
\end{align}
For the proof of \eqref{identity-3}, first, we take the derivatives of $D_{1}$. From a similar argument to ($\ref{d1_equ}$), under assumptions of the theorem,  we have
\begin{align*}
E[D_{1}|\mathbf{Q}(\mathbf{Z})]=E[\mathbbm{1}(d_{1}(\mathbf{V},\mathbf{q})=1)].
\end{align*}
Hence, as in \eqref{D_1_original_step1}, \eqref{diff_order_1-0-2}, \eqref{diff_order_1-2-0} and \eqref{D1_Step2}, we have
\begin{align}
    &\left.\frac{\partial}{\partial q_{1}}E[D_{1}|Q_{1}(\mathbf{Z})=q_{1},Q_{2}(\mathbf{Z})=q_{2}]\right|_{q_{1}=q_{1}^{*}} \notag \\
=&-\int_{0}^{q_{2}}f_{\mathbf{V}}(q_{1}^{*},v_{2})dv_{2}-\int_{q_{2}}^{1}f_{\mathbf{V}}(b(v_{2},q_{1}^{*},q_{2}),v_{2})\frac{f_{U_{1}}(F_{U_{1}}^{-1}(b(v_{2},q_{1}^{*},q_{2}))))}{f_{U_{1}}(F_{U_{1}}^{-1}(q_{1}^{*}))}dv_{2}. \label{D1_q1_Step3} \\
&\left.\frac{\partial}{\partial q_{2}}E[D_{1}|Q_{1}(\mathbf{Z})=q_{1},Q_{2}(\mathbf{Z})=q_{2}]\right|_{q_{2}=q_{2}} \notag \\
=&\int_{q_{1}}^{1}f_{\mathbf{V}}(v_{1},a(v_{1},q_{1},q_{2}))\frac{f_{U_{2}}(F_{U_{2}}^{-1}(a(v_{1},q_{1},q_{2})))}{f_{U_{2}}(F_{U_{2}}^{-1}(q_{2})))}dv_{1}. \label{D1_q2_Step3}
\end{align}
From the same argument as the one leading to \eqref{D0_outcome_q_1_derivative}, we have
\begin{equation}\label{D0_q1_Step3}
    \left.\frac{\partial}{\partial q_{1}}E[D_{0}|Q_{1}(\mathbf{Z})=q_{1},Q_{2}(\mathbf{Z})=q_{2}]\right|_{q_{1}=q_{1}^{*}}=\int_{0}^{q_{2}}f_{\mathbf{V}}(q_{1}^{*},v_{2})dv_{2}.
\end{equation}
Therefore, the required result\eqref{identity-3} follows from \eqref{D1_q1_Step3}, \eqref{D1_q2_Step3} and \eqref{D0_q1_Step3}.

From \eqref{D_1_original_step1}, \eqref{diff_order_1-0-2}, \eqref{B_q_1_new} and \eqref{identity-3}, we obtain
\begin{align*}
&\int_{0}^{q_{2}}E[G(Y_{1})|V_{1}=q_{1}^{*},V_{2}=v_{2}]f_{\mathbf{V}}(q_{1}^{*},v_{2})dv_{2} \\
=&-\left.\frac{\partial}{\partial q_{1}}E[G(Y)D_{1}|Q_{1}(\mathbf{Z})=q_{1},Q_{2}(\mathbf{Z})=q_{2}]\right|_{q_{1}=q_{1}^{*}}   \\
&+\left.\frac{\partial}{\partial q_{2}}E[G(Y)D_{1}|Q_{1}(\mathbf{Z})=q_{1}^{*},Q_{2}(\mathbf{Z})=q_{2}]\right|_{q_{2}=q_{2}} \\
&\times\frac{\left.\frac{\partial}{\partial q_{1}}E[(D_{0}+D_{1})|Q_{1}(\mathbf{Z})=q_{1},Q_{2}(\mathbf{Z})=q_{2}]\right|_{q_{1}=q_{1}^{*}}}{\left.\frac{\partial}{\partial q_{2}}E[D_{1}|Q_{1}(\mathbf{Z})=q_{1}^{*},Q_{2}(\mathbf{Z})=q_{2}]\right|_{q_{2}=q_{2}}} \\
=&-\Delta GD_{(1,0)}(1)(q_{1}^{*},q_{2})+\Delta GD_{(0,1)}(1)(q_{1}^{*},q_{2})\times\frac{\Delta D_{(1,0)}(0)(q_{1}^{*},q_{2})+\Delta D_{(1,0)}(1)(q_{1}^{*},q_{2})}{\Delta D_{(0,1)}(1)(q_{1}^{*},q_{2})}
\end{align*}
where we define
\begin{align*}
    \Delta GD_{(\ell,m)}(k)(q_{1},q_{2})&:=\left.\frac{\partial^{(\ell+m)} E[G(Y)D_{k}|Q_{1}(\mathbf{Z}),Q_{2}(\mathbf{Z})]}{\partial^{\ell} Q_{1}(\mathbf{Z})\partial^{m} Q_{2}(\mathbf{Z})}\right|_{(Q_{1}(\mathbf{Z}),Q_{2}(\mathbf{Z}))=(q_{1},q_{2})}, \\
    \Delta D_{(\ell,m)}(k)(q_{1},q_{2})&:=\left.\frac{\partial^{(\ell+m)} E[D_{k}|Q_{1}(\mathbf{Z}),Q_{2}(\mathbf{Z})]}{\partial^{\ell} Q_{1}(\mathbf{Z})\partial^{m} Q_{2}(\mathbf{Z})}\right|_{(Q_{1}(\mathbf{Z}),Q_{2}(\mathbf{Z}))=(q_{1},q_{2})}.
\end{align*}
for any $k\in\{0,1,2\}$ and $\ell,m\in\{0,1,2\}$ such that $\ell+m\leq 2$. Especially, when we evaluate $\Delta GD_{(\ell,m)}(k)(q_{1},q_{2})$ and $
    \Delta D_{(\ell,m)}(k)(q_{1},q_{2})$ at $(q_{1}^{*},q_{2}^{*})$, we denote
\begin{align*}
    \Delta GD_{(\ell,m)}^{*}(k)&:=\Delta GD_{(\ell,m)}(k)(q_{1}^{*},q_{2}^{*}), \\
    \Delta D_{(\ell,m)}^{*}(k)&:=\Delta D_{(\ell,m)}(k)(q_{1}^{*},q_{2}^{*}),
\end{align*}
respectively. Differentiating this with respect to $q_{2}$ gives
\begin{align*}
&E[G(Y_{1})|V_{1}=q_{1}^{*},V_{2}=q_{2}^{*}]f_{\mathbf{V}}(q_{1}^{*},q_{2}^{*}) \notag  \\
=&-\Delta GD^{*}_{(1,1)}(1)-\frac{\Delta GD^{*}_{(0,1)}(1)(\Delta D^{*}_{(1,0)}(0)+\Delta D^{*}_{(1,0)}(1))\Delta D^{*}_{(0,2)}(1)}{(\Delta D^{*}_{(0,1)}(1))^{2}} \notag  \\
&+ \frac{(\Delta GD^{*}_{(0,2)}(1)(\Delta D^{*}_{(1,0)}(0)+\Delta D^{*}_{(1,0)}(1))+\Delta GD^{*}_{(0,1)}(1)(\Delta D^{*}_{(1,1)}(0)+\Delta D^{*}_{(1,1)}(1)))}{\Delta D_{(0,1)}^{*}(1)}. \notag 
\end{align*}
Therefore, in conjunction with \eqref{target_1-2}, the required result follows.



For part (c), we follow the same argument as the proof of part (b), replacing  $Q_{1}(\mathbf{Z})$ with  $Q_{2}(\mathbf{Z})$ and  $Q_{2}(\mathbf{Z})$ with  $Q_{1}(\mathbf{Z})$, respectively. Then, we obtain the following equality
\begin{align*}
&\int_{0}^{q_{1}}E[G(Y_{2})|V_{1}=v_{1},V_{2}=q_{2}^{*}]f_{\mathbf{V}}(v_{1}, q_{2}^{*})dv_{1} \\
=&-\left.\frac{\partial}{\partial q_{2}}E[G(Y)D_{2}|Q_{1}(\mathbf{Z})=q_{1},Q_{2}(\mathbf{Z})=q_{2}]\right|_{q_{2}=q_{2}^{*}} \\
&+\left.\frac{\partial}{\partial q_{1}}E[G(Y)D_{2}|Q_{1}(\mathbf{Z})=q_{1},Q_{2}(\mathbf{Z})=q_{2}^{*}]\right|_{q_{1}=q_{1}}\\
&\times\frac{\left.\frac{\partial}{\partial q_{2}}E[(D_{0}+D_{2})|Q_{1}(\mathbf{Z})=q_{1},Q_{2}(\mathbf{Z})=q_{2}]\right|_{q_{2}=q_{2}^{*}}}{\left.\frac{\partial}{\partial q_{1}}E[D_{2}|Q_{1}(\mathbf{Z})=q_{1},Q_{2}(\mathbf{Z})=q_{2}^{*}]\right|_{q_{1}=q_{1}}} \\
=&-\Delta GD_{(0,1)}(2)(q_{1},q_{2}^{*})+\Delta GD_{(1,0)}(2)(q_{1},q_{2}^{*})\times\frac{\Delta D_{(0,1)}(0)(q_{1},q_{2}^{*})+\Delta D_{(0,1)}(2)(q_{1},q_{2}^{*})}{\Delta D_{(1,0)}(2)(q_{1},q_{2}^{*})}.
\end{align*}
Taking its derivative with respect to $q_{1}$ gives
\begin{align*}
&E[G(Y_{2})|V_{1}=q_{1}^{*},V_{2}=q_{2}^{*}]f_{\mathbf{V}}(q_1^{*},q_{2}^{*}) \\
=&-\Delta GD^{*}_{(1,1)}(2)-\frac{\Delta GD^{*}_{(1,0)}(2)(\Delta D^{*}_{(0,1)}(0)+\Delta D^{*}_{(0,1)}(2))\Delta D^{*}_{(2,0)}(2)}{(\Delta D^{*}_{(1,0)}(2))^{2}} \\
&+\frac{\left(\Delta GD^{*}_{(2,0)}(2)(\Delta D^{*}_{(0,1)}(0)+\Delta D^{*}_{(0,1)}(2))+\Delta GD^{*}_{(1,0)}(2)(\Delta D^{*}_{(1,1)}(0)+\Delta D^{*}_{(1,1)}(2))\right)}{\Delta D^{*}_{(1,0)}(2)}.
\end{align*}
In conjunction with \eqref{target_1-2}, the required result follows.
    
\end{proof}

\subsection{Proof of Theorem \ref{the_idenq}}
\begin{proof}[Proof of Theorem \ref{the_idenq}]
By definition,
\begin{equation*}
H(\mathbf{Z})=\int g(v_{1},v_{2};Q_{1}(\mathbf{Z}),Q_{2}(\mathbf{Z}))f_{\mathbf{V}}(v_{1},v_{2})d\mathbf{v},
\end{equation*}
where $g(v_{1},v_{2};Q_{1}(\mathbf{Z}),Q_{2}(\mathbf{Z}))=\mathbbm{1}\{v_{1}<Q_{1}(\mathbf{Z})\}\mathbbm{1}\{v_{2}<Q_{2}(\mathbf{Z})\}$. Because we define
\begin{equation*}
Q_{2}(\mathbf{Z})=F_{U_{2}}(R_{2}(\mathbf{Z})), 
\end{equation*}
by Assumption \ref{ass_iden}, we obtain
\begin{equation*}
\lim_{z^{[\ell_{2}]}\rightarrow a^{[\ell_{2}]}}Q_{2}(\mathbf{Z})=1.
\end{equation*}
Therefore, as $z^{[\ell_{2}]}\rightarrow a^{[\ell_{2}]}$,
\begin{equation*}
    g(v_{1},v_{2};Q_{1}(\mathbf{Z}),Q_{2}(\mathbf{Z}))f_{\mathbf{V}}(\mathbf{v})\rightarrow\mathbbm{1}(v_{1}<Q_{1}(\mathbf{Z}))f_{\mathbf{V}}(\mathbf{v}),\ \ \ a.s.
\end{equation*}
Hence, it follows from the dominated convergence theorem (DCT) and Fubini's theorem that
\begin{align*}
\lim_{z^{[\ell_{2}]}\rightarrow a^{[\ell_{2}]}}H(\mathbf{Z})=&\lim_{z^{[\ell_{2}]}\rightarrow a^{[\ell_{2}]}}\int g(v_{1},v_{2};Q_{1}(\mathbf{Z}),Q_{2}(\mathbf{Z}))f_{\mathbf{V}}(\mathbf{v})d\mathbf{v} \\
=&\int\mathbbm{1}(v_{1}<Q_{1}(\mathbf{Z}))f_{\mathbf{V}}(\mathbf{v})d\mathbf{v} \\
=&\int^{1}_{0}\mathbbm{1}(v_{1}<Q_{1}(\mathbf{Z}))dv_{1}=Q_{1}(\mathbf{Z}),
\end{align*}
giving the stated result for $Q_{1}(\mathbf{Z})$.

For $Q_{2}(\mathbf{Z})$, similar to the proof for $Q_{1}(\mathbf{Z})$, we have
\begin{equation*}
Q_{1}(\mathbf{Z})=F_{U_{1}}(R_{1}(\mathbf{Z})).
\end{equation*}
From Assumption \ref{ass_iden}, we obtain
\begin{equation*}
\lim_{z^{[\ell_{1}]}\rightarrow a^{[\ell_{1}]}}Q_{1}(\mathbf{Z})=1.
\end{equation*}
Therefore, as $z^{[\ell_{1}]}\rightarrow a^{[\ell_{1}]}$,
\begin{equation*}
    g(v_{1},v_{2};Q_{1}(\mathbf{Z}),Q_{2}(\mathbf{Z}))f_{\mathbf{V}}(\mathbf{v})\rightarrow\mathbbm{1}(v_{2}<Q_{2}(\mathbf{Z}))f_{\mathbf{V}}(\mathbf{v}),\ \ \ a.s.
\end{equation*}
Hence, it follows from the DCT and Fubini's theorem that:
\begin{align*}
\lim_{z^{[\ell_{1}]}\rightarrow a^{[\ell_{1}]}}H(\mathbf{Z})=&\lim_{z^{[\ell_{0}]}\rightarrow a^{[\ell_{0}]}}\int g(v_{1},v_{2};Q_{1}(\mathbf{Z}),Q_{2}(\mathbf{Z}))f_{\mathbf{V}}(\mathbf{v})d\mathbf{v} \\
=&\int\mathbbm{1}(v_{2}<Q_{2}(\mathbf{Z}))f_{\mathbf{V}}(\mathbf{v})d\mathbf{v} \\
=&\int^{1}_{0}\mathbbm{1}(v_{2}<Q_{2}(\mathbf{Z}))dv_{2}=Q_{2}(\mathbf{Z}).
\end{align*}

\end{proof}

\subsection{Proof of Theorem \ref{the_iden_gen}}


\begin{proof}[Proof of Theorem \ref{the_iden_gen}]

For convenience, let $\mathbf{a}_{-i}$ denote the vector that removes $a_{i}$ from the original vector, namely
$\mathbf{a}_{-i}=(a_{1},\cdots, a_{i-1}, a_{i+1}, \cdots, a_{n})$. We first show that
\begin{equation}\label{target_k}
E[G(Y_{k})|\mathbf{V}=\mathbf{q}^{*}]=\left.\frac{\partial E[G(Y)D_{k}|\mathbf{Q}(\mathbf{Z})]}{\partial \mathbf{Q}(\mathbf{Z})}\right|_{\mathbf{Q(Z)}=\mathbf{q}^{*}}\left/\frac{\partial E[D_{k}|\mathbf{Q}(\mathbf{Z})]}{\partial \mathbf{Q}(\mathbf{Z})}\right|_{\mathbf{Q(Z)}=\mathbf{q}^{*}}.
\end{equation}
Let $\mathbf{Q}(\mathbf{Z})$ and $\mathbf{V}$ denote a vector of thresholds $(Q_{0}(\mathbf{Z}),\cdots,Q_{k-1}(\mathbf{Z}),Q_{k+1}(\mathbf{Z})\cdots, \\ Q_{K-1}(\mathbf{Z}))^{'}$ and heterogeneity $\mathbf{V}=(V_{0},\cdots,V_{k-1},V_{k+1},\cdots,V_{K-1})^{'}$. Define $d_{k}(\mathbf{V},\mathbf{Q(Z)})$ as the indicator for treatment $k$, i.e. $d_{k}(\mathbf{V},\mathbf{Q(Z)}):=\prod_{i\in\mathcal{K}\backslash \{k\}}\mathbbm{1}\{V_{i}<Q_{i}(\mathbf{Z})\}$. Set $\mathbf{q}=(q_{0},\cdots,q_{k-1},q_{k+1},\cdots,q_{K-1})^{'}$. Under assumptions of the theorem, for any $\mathbf{q}$ in the range of $\mathbf{Q(Z)}$, a similar procedure to \eqref{d1_equ} gives
\begin{align}
E[G(Y)D_{k}|\mathbf{Q(Z)}=\mathbf{q}] 
=E[E[G(Y_{k})|\mathbf{V}]\mathbbm{1}(d_{k}(\mathbf{V},\mathbf{q})=1)].  \label{dk_equ_gen}
\end{align}
 Hence, for any $\mathbf{q}$ in the range of $\mathbf{Q(Z)}$, we have 
\begin{equation}\label{Dk_int_gen}
E[G(Y)D_{k}|\mathbf{Q({Z})}=\mathbf{q}]=\int E[G(Y_{k})|\mathbf{V}=\mathbf{v}]g_{\mathcal{K},k}(\mathbf{v};\mathbf{q})f_{\mathbf{V}}(\mathbf{v})d\mathbf{v},
\end{equation}
where $\mathbf{v}=(v_{0},\cdots,v_{k-1},v_{k+1},\cdots,v_{K-1})^{'}$ and $g_{\mathcal{K},k}(\mathbf{v};\mathbf{q}):=\prod_{i\in\mathcal{K}\backslash \{k\}}\mathbbm{1}\{v_{i}<q_{i}\}$.


From Fubini's theorem, the RHS of \eqref{Dk_int_gen} is written as 
\begin{align*}
\int\int_{0}^{q_{0}} E[G(Y_{k})|\mathbf{V}=\mathbf{v}]f_{\mathbf{V}}(\mathbf{v})dv_{0} \prod_{i\in\mathcal{K}\backslash \{0,k\}}\mathbbm{1}\{v_{i}<q_{i}\} d\mathbf{v}_{-0}.
\end{align*}
It follows from Assumption \ref{new_assumption_dif_multiple} \eqref{continuity_v_multiple} and the Leibniz integral rule that, for any $\mathbf{v}_{-0}\in(0,1)^{K-2}$
\begin{equation*}
    \left.\frac{\partial}{\partial q_{0}}\int_{0}^{q_{0}} E[G(Y_{k})|\mathbf{V}=\mathbf{v}]f_{\mathbf{V}}(\mathbf{v})dv_{0} \right|_{q_{0}=q_{0}^{*}}=E[G(Y_{k})|V_{0}=q_{0}^{*},\mathbf{V}_{-0}=\mathbf{v}_{-0}]f_{\mathbf{V}}(q_{0}^{*},\mathbf{v}_{-0}).
\end{equation*}
Furthermore, because $\sup_{\mathbf{v}\in(0,1)^{K-1}}E[|G(Y_{k})||\mathbf{V}=\mathbf{v}]$ is finite from Assumption \ref{new_assumption_dif_multiple} \eqref{bounded_moment_multiple}
and $f_{\mathbf{V}}(q_{0}^{*},\mathbf{v}_{-0})$ is integrable with respect to $\mathbf{v}_{-0}$, we can exchange differentiation and integral to obtain
\begin{align}\label{Dk_outcome_q_0_derivative}
    &\left.\frac{\partial}{\partial q_{0}}E[G(Y)D_{k}|\mathbf{Q({Z})}=\mathbf{q}] \right|_{q_{0}=q_{0}^{*}} \notag \\
    =&\int E[G(Y_{k})|V_{0}=q_{0}^{*},\mathbf{V}_{-0}=\mathbf{v}_{-0}]f_{\mathbf{V}}(q_{0}^{*},\mathbf{v}_{-0}) \prod_{i\in\mathcal{K}\backslash \{0,k\}}\mathbbm{1}\{v_{i}<q_{i}\} d\mathbf{v}_{-0}.
\end{align}
By iterating the above process to each element in $\mathbf{v}_{-0}$, we obtain
\begin{equation}\label{Dk_outcome_result}
\left.\frac{\partial E[G(Y)D_{k}|\mathbf{Q}(\mathbf{Z})]}{\partial \mathbf{Q}(\mathbf{Z})}\right|_{\mathbf{Q(Z)}=\mathbf{q}^{*}}=E[G(Y_{k})|\mathbf{V}=\mathbf{q}^{*}]f_{\mathbf{V}}(\mathbf{q}^{*}).
\end{equation}
As an extension of the three valued case, we can also obtain  
\begin{equation}
         f_{\mathbf{V}}(\mathbf{q}^{*})=\left.\frac{\partial E[D_{k}|\mathbf{Q}(\mathbf{Z})]}{\partial \mathbf{Q}(\mathbf{Z})}\right|_{\mathbf{Q(Z)}=\mathbf{q}^{*}}. \label{target_density_multiple}
\end{equation}
Therefore, the required result holds.

We move on the proof of the identification for $E[G(Y_{j})|\mathbf{V}=\mathbf{q}^{*}]$. Let $d_{j}(\mathbf{V},\mathbf{Q}(\mathbf{Z}))$ denote the indicator for treatment $j$, i.e. $d_{j}(\mathbf{V},\mathbf{Q}(\mathbf{Z}))=\mathbbm{1}\{Q_{j}(\mathbf{Z})\leq V_{j}\} \\
\times \prod_{i\in\mathcal{K}\backslash \{j,k\}}\mathbbm{1}\left\{V_{i}<F_{\widetilde{U}_{i,k}}(F^{-1}_{\widetilde{U}_{j,k}}(V_{j})-F^{-1}_{\widetilde{U}_{j,k}}(Q_{j}(\mathbf{Z}))+F^{-1}_{\widetilde{U}_{i,k}}(Q_{i}(\mathbf{Z})))\right\}$. From a similar argument to \eqref{dk_equ_gen}, under assumptions of the theorem, for any $\mathbf{q}$ in the range of $\mathbf{Q(Z)}$,  we obtain
\begin{equation}\label{Dj_int_gen}
E[G(Y)D_{j}|\mathbf{Q({Z})}=\mathbf{q}]=\int E[G(Y_{j})|\mathbf{V}=\mathbf{v}]\ell_{\mathcal{K},k,j}(\mathbf{v};\mathbf{q})f_{\mathbf{V}}(\mathbf{v})d\mathbf{v},
\end{equation}
where we define
\begin{equation*}
    \ell_{\mathcal{K},k,j}(\mathbf{v};\mathbf{q}):=\mathbbm{1}\{q_{j}\leq v_{j}\}\times\prod_{i\in\mathcal{K}\backslash \{j,k\}}\mathbbm{1}\left\{v_{i}<F_{\widetilde{U}_{i,k}}(F^{-1}_{\widetilde{U}_{j,k}}(v_{j})-F^{-1}_{\widetilde{U}_{j,k}}(q_{j})+F^{-1}_{\widetilde{U}_{i,k}}(q_{i}))\right\}.
\end{equation*}

Let $\mathcal{V}_{K-1}:=(0,1)^{K-1}$ denote the support of $\mathbf{V}$, and let $\mathcal{V}_{K-1}\{a\succ b\}$ denote the support of $\mathbf{V}$ with the first ranked treatment $a$ and the second ranked treatment $b$. In the following, let $q_{i}$ denote arbitrary points of neighborhoods of $q_{i}^{*}$.

We decompose the domain of integration in \eqref{Dj_int_gen} into $K-1$ areas ranging from  $\mathcal{V}\{j\succ 0\}$ to  $\mathcal{V}\{j\succ K-1\}$,
\begin{align}
   &E[G(Y)D_{j}|\mathbf{Q({Z})}=\mathbf{q}]=\sum_{i=0, i\neq j,k}^{K-1}A_{i}(\mathbf{q})\label{D_j_original_step1} 
\end{align}
where
\begin{align*}
A_{k}(\mathbf{q})&:=\int E[G(Y_{j})|\mathbf{V}=\mathbf{v}]f_{\mathbf{V}}(\mathbf{v})\mathbbm{1}\{v_{j}\geq q_{j}\}\prod_{i\in\mathcal{K}\backslash \{k,j\}}\mathbbm{1}\{v_{i}<q_{i}\}d\mathbf{v} , \\
A_{i}(\mathbf{q})&:=\int E[G(Y_{j})|\mathbf{V}=\mathbf{v}]f_{\mathbf{V}}(\mathbf{v})\mathbbm{1}\left\{c^{i,j}_{k}(v_{i},q_{i},q_{j})\leq v_{j}\right\} \notag \\
    \times &\mathbbm{1}\{v_{i}\geq q_{i}\}\prod_{\ell\in\mathcal{K}\backslash \{k, i, j\}}\mathbbm{1}\left\{v_{\ell}<c^{i,\ell}_{k}(v_{i},q_{i},q_{\ell})\right\}d\mathbf{v} \quad \text{for each $i\in \mathcal{K}^{\backslash\{k,j\}}$}\label{order_j-other}
, \\
c^{i,\ell}_{k}(v_{i},q_{i},q_{\ell})&:=F_{\tilde{U}_{\ell,k}}(F_{\tilde{U}_{i,k}}^{-1}(v_{i})-F_{\tilde{U}_{i,k}}^{-1}(q_{i})+F_{\tilde{U}_{\ell,k}}^{-1}(q_{\ell}))\quad  \text{for each $i\in \mathcal{K}^{\backslash\{k,j\}}$ and $\ell\in\mathcal{K}^{\backslash\{k,i\}} $}
\end{align*}
For each $\ell\in\mathcal{K}^{\backslash\{j\}} $,
$ A_{\ell}(\mathbf{q})$ correspond to $\mathcal{V}\{j\succ\ell\}$, respectively. 

First, we examine the effect of a marginal change in $Q_{j}(\mathbf{Z})$ on $A_{k}(\mathbf{q})$. Because we can exchange the order of differentiation and integration as in the case of treatment $k$, we obtain 
\begin{equation}\label{diff_order_j-k}
\begin{split}
     &\left.\frac{\partial}{\partial q_{j}}\int E[G(Y_{j})|\mathbf{V}=\mathbf{v}]f_{\mathbf{V}}(\mathbf{v})\mathbbm{1}\{v_{j}\geq q_{j}\}\prod_{i\in\mathcal{K}\backslash \{k,j\}}\mathbbm{1}\{v_{i}<q_{i}\}d\mathbf{v}\right|_{q_{j}=q_{j}^{*}} \\
    =&-\int E[G(Y_{j})|V_{j}=q_{j}^{*},\mathbf{V}_{-j}=\mathbf{v}_{-j}]f_{\mathbf{V}}(V_{j}=q_{j}^{*},\mathbf{V}_{-j}=\mathbf{v}_{-j})\prod_{i\in\mathcal{K}\backslash \{k,j\}}\mathbbm{1}\{v_{i}<q_{i}\}d\mathbf{v}_{-j}.
\end{split}
\end{equation}

Second, we examine the marginal change of $Q_{j}(\mathbf{Z})$ on $A_{i}(\mathbf{q})$ for each $i\in \mathcal{K}^{\backslash\{k,j\}}$. It holds from Assumption \ref{new_assumption_dif_multiple} \eqref{diff_q_multiple}, \eqref{continuity_v_multiple} and  the Leibniz integral rule that, for any $\mathbf{v}_{-j}\in(0,1)^{K-1}$
\begin{align*}
    &\frac{\partial}{\partial q_{j}}\left.\int_{c^{i,j}_{k}(v_{i},q_{i},q_{j})}^{1}E[G(Y_{j})|\mathbf{V}=\mathbf{v}]f_{\mathbf{V}}(\mathbf{v})\mathbbm{1}\{v_{i}\geq q_{i}\}\times \prod_{\ell\in\mathcal{K}\backslash \{k, i, j\}}\mathbbm{1}\left\{v_{\ell}<c^{i,\ell}_{k}(v_{i},q_{i},q_{\ell})\right\}d\mathbf{v}\right|_{q_{j}=q_{j}^{*}} \\
    =&-E[G(Y_{j})|V_{j}=c^{i,j}_{k}(v_{i},q_{i},q_{j}^{*}),\mathbf{V}_{-j}=\mathbf{v}_{-j}]f_{\mathbf{V}}(V_{j}=c^{i,j}_{k}(v_{i},q_{i},q_{j}^{*}),\mathbf{V}_{-j}=\mathbf{v}_{-j}) \\
    \times&\frac{f_{\tilde{U}_{j,k}}(F_{\tilde{U}_{j,k}}^{-1}(c^{i,j}_{k}(v_{i},q_{i},q_{j}^{*})))}{f_{\tilde{U}_{j,k}}(F_{\tilde{U}_{j,k}}^{-1}(q_{j}^{*}))}  \mathbbm{1}\{v_{i}\geq q_{i}\}\prod_{\ell\in\mathcal{K}\backslash \{k, i, j\}}\mathbbm{1}\left\{v_{\ell}<c^{i,\ell}_{k}(v_{i},q_{i},q_{\ell})\right\}
\end{align*}
Through the argument of the change of variables, we have
\begin{equation*}
    f_{\mathbf{V}}(v_{0},\cdots,v_{K-1})=\frac{f_{\tilde{U}_{0,k},\cdots, \tilde{U}_{K-1,k}}(F_{\tilde{U}_{0,k}}^{-1}(v_{0}),\cdots, F_{\tilde{U}_{K-1,k}}^{-1}(v_{K-1}))}{\prod_{\ell\in\mathcal{K}\backslash\{k\}}f_{\tilde{U}_{\ell,k}}(F_{\tilde{U}_{\ell,k}}^{-1}(v_{\ell}))}.
\end{equation*}
Because 
\begin{align*}
    \sup_{\mathbf{v}\in(0,1)^{(K-1)}}&E[|G(Y_{i})||\mathbf{V}=\mathbf{v}]<\infty\quad \\
    \text{and}\quad \sup_{(\tilde{u}_{0,k},\cdots, \tilde{u}_{K-1,k})\in\mathbb{R}^{K-1}}&\frac{f_{\tilde{U}_{0,k},\cdots, \tilde{U}_{K-1,k}}(\tilde{u}_{0,k},\cdots, \tilde{u}_{K-1,k})}{\prod_{\ell\in\mathcal{K}\backslash\{k, j\}}f_{\tilde{U}_{\ell,k}}(\tilde{u}_{\ell,k})}<\infty
\end{align*}
hold, we can exchange the order of differentiation and integration. Therefore, we obtain 
\begin{equation}\label{diff_order_j_other}
\begin{split}
     &\left.\frac{\partial}{\partial q_{j}}A_{i}(\mathbf{q})\right|_{q_{j}=q_{j}^{*}} \\
=-&\int E[G(Y_{j})|V_{j}=c^{i,j}_{k}(v_{i},q_{i},q_{j}^{*}),\mathbf{V}_{-j}=\mathbf{v}_{-j}]
f_{\mathbf{V}}(V_{j}=c^{i,j}_{k}(v_{i},q_{i},q_{j}^{*}),\mathbf{V}_{-j}=\mathbf{v}_{-j}) \\
    \times&\frac{f_{\tilde{U}_{j,k}}(F_{\tilde{U}_{j,k}}^{-1}(c^{i,j}_{k}(v_{i},q_{i},q_{j}^{*})))}{f_{\tilde{U}_{j,k}}(F_{\tilde{U}_{j,k}}^{-1}(q_{j}^{*}))}  \mathbbm{1}\{v_{i}\geq q_{i}\}\prod_{\ell\in\mathcal{K}\backslash \{k, i, j\}}\mathbbm{1}\left\{v_{\ell}<c^{i,\ell}_{k}(v_{i},q_{i},q_{\ell})\right\}d\mathbf{v}_{-j}
\end{split}
\end{equation}


We derive an alternate expansion of the RHS of \eqref{diff_order_j_other}. First, we use the change of variables.When we set $v_{i}=c^{j,i}_{k}(v_{j},q_{j}^{*},q_{i})$, by definition, we obtain $v_{j}=c^{i,j}_{k}(v_{i},q_{i},q_{j}^{*})$. As in the proof of Theorem \ref{the_iden}, it holds from \eqref{diff_order_j_other} that 
\begin{align}
&\int E[G(Y_{j})|V_{j}=c^{i,j}_{k}(v_{i},q_{i},q_{j}^{*}), \mathbf{V}_{-j}=\mathbf{v}_{-j}]f_{\mathbf{V}}( c^{i,j}_{k}(v_{i},q_{i},q_{j}^{*}), \mathbf{v}_{-j})f_{\tilde{U}_{j,k}}(F^{-1}_{\tilde{U}_{j,k}}(c^{i,j}_{k}(v_{i},q_{i},q_{j}^{*}))) \notag \\
    \times &\mathbbm{1}\{ v_{i}\geq q_{i}\}\prod_{\ell\in\mathcal{K}\backslash \{k, i, j\}}\mathbbm{1}\left\{v_{\ell}<c^{i,\ell}_{k}(v_{i},q_{i},q_{\ell})\right\}d\mathbf{v}_{-j} \notag \\
    =    &\int E[G(Y_{j})|V_{i}=c^{j,i}_{k}(v_{j},q_{j}^{*},q_{i}), \mathbf{V}_{-i}=\mathbf{v}_{-(i,j)}]f_{\mathbf{V}}(c^{j,i}_{k}(v_{j},q_{j}^{*},q_{i}), \mathbf{v}_{-i}) \notag \\
    \times &f_{\tilde{U}_{i,k}}(F_{\tilde{U}_{i,k}}^{-1}(c^{j,i}_{k}(v_{j},q_{j}^{*},q_{i})))\mathbbm{1}\{v_{j}\geq q_{j}^{*}\} \prod_{\ell\in\mathcal{K}\backslash \{k, i, j\}}\mathbbm{1}\left\{v_{\ell}<c^{j,\ell}_{k}(v_{j},q_{j}^{*},q_{\ell})\right\}d\mathbf{v}_{-i}
    \label{Dj_Step3-3}
\end{align}
Second, we express \eqref{Dj_Step3-3} as a function of  a partial derivative of $E[G(Y)D_{j}|\mathbf{Q}(\mathbf{Z})=\mathbf{q}]$ with respect to $Q_{i}(\mathbf{Z})$.  It holds from  Assumption \ref{new_assumption_dif_multiple} \eqref{continuity_v_multiple} and the Leibniz integral rule that, for any $v_{i}\in(0,1)$
\begin{align*}
    &\left.\frac{\partial}{\partial q_{i}}\int_{0}^{c^{j,i}_{k}(v_{j},q_{j},q_{i})}E[G(Y_{j})|V_{i}=v_{i},\mathbf{V}_{-i}=\mathbf{v}_{-i}]f_{\mathbf{V}}(V_{i}=v_{i},\mathbf{V}_{-i}=\mathbf{v}_{-i})dv_{i}\right|_{q_{i}=q_{i}} \notag \\
    =&E[G(Y_{j})|V_{i}=c^{j,i}_{k}(v_{j},q_{j},q_{i}),\mathbf{V}_{-i}=\mathbf{v}_{-i}]f_{\mathbf{V}}(c^{j,i}_{k}(v_{j},q_{j},q_{i}), \mathbf{v}_{-i})\frac{f_{\tilde{U}_{i,k}}(F_{\tilde{U}_{i,k}}^{-1}(c^{j,i}_{k}(v_{j},q_{j},q_{i})))}{f_{\tilde{U}_{i,k}}(F_{\tilde{U}_{i,k}}^{-1}(q_{i}))} 
\end{align*}
As in the case of \eqref{diff_order_j_other}, we can exchange the order of differentiation and integration. Hence, differentiating \eqref{Dj_int_gen} with respect to $q_{i}$ gives
\begin{align}
    &\left.\frac{\partial}{\partial q_{i}}E[G(Y)D_{j}|\mathbf{Q}(\mathbf{Z})=\mathbf{q}]\right|_{(\mathbf{q}_{-j}, q_{j})=(\mathbf{q}_{-j}, q_{j}^{*})} \notag \\
    =&\int E[G(Y_{j})|V_{i}=c^{j,i}_{k}(v_{j},q_{j},q_{i}),\mathbf{V}_{-i}=\mathbf{v}_{-i}]f_{\mathbf{V}}(c^{j,i}_{k}(v_{j},q_{j}^{*},q_{i}), \mathbf{v}_{-i})\frac{f_{\tilde{U}_{i,k}}(F_{\tilde{U}_{i,k}}^{-1}(c^{j,i}_{k}(v_{j},q_{j},q_{i})))}{f_{\tilde{U}_{i,k}}(F_{\tilde{U}_{i,k}}^{-1}(q_{i}))}  \notag \\
    \times&\mathbbm{1}\{v_{j}\geq q_{j}^{*}\}  \prod_{\ell\in\mathcal{K}\backslash \{k, i, j\}}\mathbbm{1}\left\{v_{\ell}<c^{j,\ell}_{k}(v_{j},q_{j}^{*},q_{\ell})\right\}d\mathbf{v}_{-i}.
    \label{Dj_Step2}
\end{align}
In conjunction with \eqref{diff_order_j_other} and \eqref{Dj_Step3-3}, we have
\begin{equation}\label{A_i_q_j_new}
    \left.\frac{\partial}{\partial q_{j}}A_{i}(\mathbf{q})\right|_{q_{j}=q_{j}^{*}} =-\left.\frac{\partial}{\partial q_{i}}E[G(Y)D_{j}|\mathbf{Q}(\mathbf{Z})=\mathbf{q}]\right|_{(\mathbf{q}_{-j}, q_{j})=(\mathbf{q}_{-j}, q_{j}^{*})}\times \frac{f_{\tilde{U}_{i,k}}(F_{\tilde{U}_{i,k}}^{-1}(q_{i}))}{f_{\tilde{U}_{j,k}}(F_{\tilde{U}_{j,k}}^{-1}(q_{j}^{*}))}
\end{equation}
Therefore, it holds from \eqref{diff_order_j-k} and \eqref{A_i_q_j_new} that
\begin{align}
    &\left.\frac{\partial}{\partial q_{j}}E[G(Y)D_{j}|\mathbf{Q}(\mathbf{Z})=\mathbf{q}]\right|_{(\mathbf{q}_{-j}, q_{j})=(\mathbf{q}_{-j}, q_{j}^{*})} \notag \\
    +&\int E[G(Y_{j})|V_{j}=q_{j}^{*},\mathbf{V}_{-j}=\mathbf{v}_{-j}]f_{\mathbf{V}}(V_{j}=q_{j}^{*},\mathbf{V}_{-j}=\mathbf{v}_{-j})\prod_{i\in\mathcal{K}\backslash \{k,j\}}\mathbbm{1}\{v_{i}<q_{i}\}d\mathbf{v}_{-j}\notag \\
  =&-\sum_{i\neq k,j}^{K-1}\left(\frac{f_{\tilde{U}_{i,k}}(F_{\tilde{U}_{i,k}}^{-1}(q_{i}))}{f_{\tilde{U}_{j,k}}(F_{\tilde{U}_{j,k}}^{-1}(q_{j}^{*}))}\times\left.\frac{\partial}{\partial q_{i}}E[G(Y)D_{j}|\mathbf{Q}(\mathbf{Z})=\mathbf{q}]\right|_{(\mathbf{q}_{-j}, q_{j})=(\mathbf{q}_{-j}, q_{j}^{*})} \right)\notag \\
  =&-\frac{f_{\tilde{U}_{b^{*},k}}(F_{\tilde{U}_{b^{*},k}}^{-1}(q_{b^{*}}))}{f_{\tilde{U}_{j,k}}(F_{\tilde{U}_{j,k}}^{-1}(q_{j}^{*}))}\sum_{i\neq k,j}^{K-1}\left(\frac{f_{\tilde{U}_{i,k}}(F_{\tilde{U}_{i,k}}^{-1}(q_{i}))}{f_{\tilde{U}_{b^{*},k}}(F_{\tilde{U}_{b^{*},k}}^{-1}(q_{b^{*}}))}\times\left.\frac{\partial}{\partial q_{i}}E[G(Y)D_{j}|\mathbf{Q}(\mathbf{Z})=\mathbf{q}]\right|_{(\mathbf{q}_{-j}, q_{j})=(\mathbf{q}_{-j}, q_{j}^{*})} \right)\label{nominator_gen}
\end{align}

We consider derivatives of $D_{j}$ and $D_{k}$. For the derivative of $D_{k}$, we have
\begin{equation}\label{Dk_q1_Step3}
    \left.\frac{\partial}{\partial q_{j}}E[D_{k}|\mathbf{Q}(\mathbf{Z})=\mathbf{q}]\right|_{(\mathbf{q}_{-j}, q_{j})=(\mathbf{q}_{-j}, q_{j}^{*})} 
=\int f_{\mathbf{V}}(q_{j}^{*},\mathbf{v}_{-j})\prod_{i\in\mathcal{K}\backslash \{k,j\}}\mathbbm{1}\{v_{i}<q_{i}\}d\mathbf{v}_{-j} 
\end{equation}
For the derivative of $D_{j}$, from a similar argument to the one leading to  \eqref{diff_order_j-k} and \eqref{diff_order_j_other}, we have
\begin{align}
    &\left.\frac{\partial}{\partial q_{j}}E[D_{j}|\mathbf{Q}(\mathbf{Z})=\mathbf{q}]\right|_{(\mathbf{q}_{-j}, q_{j})=(\mathbf{q}_{-j}, q_{j}^{*})} \notag \\
=-&\int f_{\mathbf{V}}(q_{j}^{*},\mathbf{v}_{-j})\prod_{i\in\mathcal{K}\backslash \{k,j\}}\mathbbm{1}\{v_{i}<q_{i}\}d\mathbf{v}_{-j} \notag \\
-&\sum_{i\neq k,j}^{K-1}\int f_{\mathbf{V}}(c^{i,j}_{k}(v_{i},q_{i},q_{j}^{*}),\mathbf{v}_{-j})\frac{f_{\tilde{U}_{j,k}}(F_{\tilde{U}_{j,k}}^{-1}(c^{i,j}_{k}(v_{i},q_{i},q_{j}^{*})))}{f_{\tilde{U}_{j,k}}(F_{\tilde{U}_{j,k}}^{-1}(q_{j}^{*}))}\mathbbm{1}\{v_{i}\geq q_{i}\} \notag \\
\times &\prod_{\ell\in\mathcal{K}\backslash \{k, i, j\}}\mathbbm{1}\left\{v_{\ell}<c^{i,\ell}_{k} (v_{i},q_{i},q_{\ell})\right\}d\mathbf{v}_{-j}. \label{Dj_q1_Step3} 
\end{align}
Moreover, for any $i\in \mathcal{K}^{\backslash\{k,j\}} $, through the argument of the change of variables used in \eqref{Dj_Step3-3}, we also have
\begin{align}
&\left.\frac{\partial}{\partial q_{i}}E[D_{j}|\mathbf{Q}(\mathbf{Z})=\mathbf{q}]\right|_{(\mathbf{q}_{-j}, q_{j})=(\mathbf{q}_{-j}, q_{j}^{*})} \notag \\
=&\int f_{\mathbf{V}}(c^{i,j}_{k}(v_{i},q_{i},q_{j}^{*}),\mathbf{v}_{-j})\frac{f_{\tilde{U}_{j,k}}(F_{\tilde{U}_{j,k}}^{-1}(c^{i,j}_{k}(v_{i},q_{i},q_{j}^{*})))}{f_{\tilde{U}_{i,k}}(F_{\tilde{U}_{i,k}}^{-1}(q_{i}))}\mathbbm{1}\{v_{i}\geq q_{i}\} \notag \\
\times &\prod_{\ell\in\mathcal{K}\backslash \{k, i, j\}}\mathbbm{1}\left\{v_{\ell}<c^{i,\ell}_{k} (v_{i},q_{i},q_{\ell})\right\}d\mathbf{v}_{-j}. \label{Dj_q2_Step3}
\end{align}
Therefore, it holds from \eqref{Dk_q1_Step3}, \eqref{Dj_q1_Step3} and \eqref{Dj_q2_Step3} that we have
\begin{align}
    &\left.\frac{\partial}{\partial q_{j}}E[(D_{j}+D_{k})|\mathbf{Q}(\mathbf{Z})=\mathbf{q}]\right|_{(\mathbf{q}_{-j}, q_{j})=(\mathbf{q}_{-j}, q_{j}^{*})} \notag \\
    =&-\sum_{i\neq k,j}^{K-1}\left(\frac{f_{\tilde{U}_{i,k}}(F_{\tilde{U}_{i,k}}^{-1}(q_{i}))}{f_{\tilde{U}_{j,k}}(F_{\tilde{U}_{j,k}}^{-1}(q^{*}_{j}))}\times\left.\frac{\partial}{\partial q_{i}}E[D_{j}|\mathbf{Q}(\mathbf{Z})=\mathbf{q}]\right|_{(\mathbf{q}_{-j}, q_{j})=(\mathbf{q}_{-j}, q_{j}^{*})} \right) \notag \\
    =&-\frac{f_{\tilde{U}_{b^{*},k}}(F_{\tilde{U}_{b^{*},k}}^{-1}(q_{b^{*}}))}{f_{\tilde{U}_{j,k}}(F_{\tilde{U}_{j,k}}^{-1}(q_{j}^{*}))}\sum_{i\neq k,j}^{K-1}\left(\frac{f_{\tilde{U}_{i,k}}(F_{\tilde{U}_{i,k}}^{-1}(q_{i}))}{f_{\tilde{U}_{b^{*},k}}(F_{\tilde{U}_{b^{*},k}}^{-1}(q_{b^{*}}))}  \times\left.\frac{\partial}{\partial q_{i}}E[D_{j}|\mathbf{Q}(\mathbf{Z})=\mathbf{q}]\right|_{(\mathbf{q}_{-j}, q_{j})=(\mathbf{q}_{-j}, q_{j}^{*})} \right)\label{denominator_gen}
\end{align}

Finally, we derive the required result. From \eqref{nominator_gen} and \eqref{denominator_gen}, we obtain
\begin{align*}
&\int E[G(Y_{j})|\mathbf{V}=\mathbf{v}]f_{\mathbf{V}}(\mathbf{v})\prod_{i\in\mathcal{K}\backslash \{k,j\}}\mathbbm{1}\{v_{i}<q_{i}\}d\mathbf{v}_{-j} \\
=
     &\frac{\sum_{i\neq k,j}^{K-1}\left(\frac{f_{\tilde{U}_{i,k}}(F_{\tilde{U}_{i,k}}^{-1}(q_{i}))}{f_{\tilde{U}_{b^{*},k}}(F_{\tilde{U}_{b^{*},k}}^{-1}(q_{b^{*}}))}\times\left.\frac{\partial}{\partial q_{i}}E[G(Y)D_{j}|\mathbf{Q}(\mathbf{Z})=\mathbf{q}]\right|_{(\mathbf{q}_{-j}, q_{j})=(\mathbf{q}_{-j}, q_{j}^{*})} \right)}{\sum_{i\neq k,j}^{K-1}\left(\frac{f_{\tilde{U}_{i,k}}(F_{\tilde{U}_{i,k}}^{-1}(q_{i}))}{f_{\tilde{U}_{b^{*},k}}(F_{\tilde{U}_{b^{*},k}}^{-1}(q_{b^{*}}))}\times\left.\frac{\partial}{\partial q_{i}}E[D_{j}|\mathbf{Q}(\mathbf{Z})=\mathbf{q}]\right|_{(\mathbf{q}_{-j}, q_{j})=(\mathbf{q}_{-j}, q_{j}^{*})} \right) }   \\
    \times& \left.\frac{\partial}{\partial q_{j}}E[(D_{j}+D_{k})|\mathbf{Q}(\mathbf{Z})=\mathbf{q}]\right|_{(\mathbf{q}_{-j}, q_{j})=(\mathbf{q}_{-j}, q_{j}^{*})} \\
    -&\left.\frac{\partial}{\partial q_{j}}E[G(Y)D_{j}|\mathbf{Q}(\mathbf{Z})=\mathbf{q}]\right|_{(\mathbf{q}_{-j}, q_{j})=(\mathbf{q}_{-j}, q_{j}^{*})} 
\end{align*}
Hence, it holds from the Leibniz integral rule that we have
\begin{align*}
&E[G(Y_{j})|\mathbf{V}=\mathbf{q}^{*}]f_{\mathbf{V}}(\mathbf{q}^{*}) \notag  \\
=& \frac{\partial^{K-2}}{\partial \mathbf{Q}_{-j}(\mathbf{Z})}\left(\frac{\sum_{i\neq k,j}^{K-1}\left(\frac{f_{\tilde{U}_{i,k}}(F_{\tilde{U}_{i,k}}^{-1}(q_{i}))}{f_{\tilde{U}_{b^{*},k}}(F_{\tilde{U}_{b^{*},k}}^{-1}(q_{b^{*}}))}\times\left.\frac{\partial}{\partial q_{i}}E[G(Y)D_{j}|\mathbf{Q}(\mathbf{Z})=\mathbf{q}]\right|_{(\mathbf{q}_{-j}, q_{j})=(\mathbf{q}_{-j}, q_{j}^{*})}  \right)}{\sum_{i\neq k,j}^{K-1}\left(\frac{f_{\tilde{U}_{i,k}}(F_{\tilde{U}_{i,k}}^{-1}(q_{i}))}{f_{\tilde{U}_{b^{*},k}}(F_{\tilde{U}_{b^{*},k}}^{-1}(q_{b^{*}}))}\times\left.\frac{\partial}{\partial q_{i}}E[D_{j}|\mathbf{Q}(\mathbf{Z})=\mathbf{q}]\right|_{(\mathbf{q}_{-j}, q_{j})=(\mathbf{q}_{-j}, q_{j}^{*})}  \right) } \right.  \\
    \times& \left.\left.\frac{\partial}{\partial q_{j}}E[(D_{j}+D_{k})|\mathbf{Q}(\mathbf{Z})=\mathbf{q}]\right|_{(\mathbf{q}_{-j}, q_{j})=(\mathbf{q}_{-j}, q_{j}^{*})} \right. \\
    -&\left.\left.\left.\frac{\partial}{\partial q_{j}}E[G(Y)D_{j}|\mathbf{Q}(\mathbf{Z})=\mathbf{q}]\right|_{(\mathbf{q}_{-j}, q_{j})=(\mathbf{q}_{-j}, q_{j}^{*})}\right)\right|_{\mathbf{q}=\mathbf{q}^{*}} .
\end{align*}
Therefore, in conjunction with \eqref{target_density_multiple}, the required result follows.
\end{proof}

\section{Sufficient condition of Assumption \ref{new_assumption_density_ratio}}\label{economics_66}

We give a sufficient condition for Assumption \ref{new_assumption_density_ratio} \eqref{new_assumption_density_ratio_ratios}. We require that there exist instrument variables that have a common constant marginal effect on the observed terms of each choice. Define $\mathbf{Z}^{*}$ as instrumental variables satisfying $\mathbf{Q}(\mathbf{Z}^{*})=\mathbf{q}^{*}$.
\begin{assumption}[Constant Marginal Effect]\label{new_assumption_imp}
Assume Assumption \ref{new_assumption_density_ratio} \eqref{new_assumption_density_ratio_differentiation} holds. For all choice $i\in\mathcal{K}^{\backslash \{j,k\}}$, there exists an instrumental variable $W_{i}$ such that $\tilde{R}_{i,k}(\mathbf{Z})$ is partially differentiable with respect to $W_{i}$ at $\mathbf{Z}=\mathbf{Z}^{*}$ and we have
\begin{equation*}
    \left.\frac{\partial \tilde{R}_{i,k}(\mathbf{Z})}{\partial W_{i}}\right|_{\mathbf{Z}=\mathbf{Z}^{*}}=\alpha
\end{equation*}
where $\alpha$ is non-zero constant. 
\end{assumption}
In Assumption \ref{new_assumption_imp}, each choice could have the same instrument variable $W_{j}$, i.e., it holds $W_{i}=W_{\ell}$ for $i,\ell\in\mathcal{K}^{\backslash \{j,k\}}$. We illustrate Assumption \ref{new_assumption_imp} in the setting of the semiparametric model. Let $\mathbf{Z}:=(W,\mathbf{S},\mathbf{Z}_{i})$ denote a set of instrumental variables. $(W,\mathbf{S})$ are instrumental variables common among alternatives and $\mathbf{Z}_{i}$ corresponds to instruments specific to the choice $i$. Assume the following form of $\tilde{R}_{i,k}(\mathbf{Z})$, 
\begin{align*}
    \tilde{R}_{i,k}(\mathbf{Z})=\alpha W+\tilde{\nu}_{i,k}(\mathbf{S},\mathbf{Z}_{i}) \quad \text{where} \quad \alpha\neq0 \quad \text{for any $i\in\mathcal{K}^{\backslash \{j,k\}}$}.
\end{align*}
where $\tilde{v}_{i,k}$ is an unknown function. 
In this setting, the marginal effect $W$ on $\tilde{R}_{i,k}(\mathbf{Z})$ constantly affects the utilities of all the choices compared to the choice $k$. 

Using Assumption \ref{new_assumption_imp}, we can establish the identification results stated in Assumption \ref{new_assumption_density_ratio} \eqref{new_assumption_density_ratio_ratios}.
\begin{lemma}\label{lemma_density}
Under Assumption \ref{new_assumption_imp}, Assumption \ref{new_assumption_density_ratio} \eqref{new_assumption_density_ratio_ratios} holds.
\end{lemma}

\begin{proof}[Proof of Lemma \ref{lemma_density}]

For the proof of this result, we define $LR(q_{i}^{*}, q_{b^{*}}^{*})$ as the density ratio of $\tilde{U}_{i,k}$ to $\tilde{U}_{b^{*},k}$, i.e.
\begin{equation*}
    LR(q_{i}^{*}, q_{b^{*}}^{*})=\frac{f_{\tilde{U}_{i,k}}(F_{\tilde{U}_{i,k}}^{-1}(q_{i}^{*}))}{f_{\tilde{U}_{b^{*},k}}(F_{\tilde{U}_{b^{*},k}}^{-1}(q_{b^{*}}^{*}))}.
\end{equation*}
A straightforward calculation gives
\begin{align*}
    \left.\frac{\partial  LR(q_{i}, q_{b^{*}}^{*})}{\partial q_{i}}\right|_{q_{i}=q_{i}^{*}}&=\frac{\partial f_{\tilde{U}_{i,k}}(F_{\tilde{U}_{i,k}}^{-1}(q_{i}^{*}))}{\partial \tilde{u}_{i,k}}\frac{1}{f^{2}_{\tilde{U}_{i,k}}(F_{\tilde{U}_{i,k}}^{-1}(q_{i}^{*}))}LR(q_{i}^{*}, q_{b^{*}}^{*}),\\
     \left.\frac{\partial  LR(q_{i}^{*}, q_{b^{*}})}{\partial q_{b^{*}}}\right|_{q_{b^{*}}=q_{b^{*}}^{*}}&=-\frac{\partial f_{\tilde{U}_{b^{*},k}}(F_{\tilde{U}_{b^{*},k}}^{-1}(q_{b^{*}}^{*}))}{\partial \tilde{u}_{b^{*},k}}\frac{1}{f^{2}_{\tilde{U}_{b^{*},k}}(F_{\tilde{U}_{b^{*},k}}^{-1}(q_{b^{*}}^{*}))}LR(q_{i}^{*}, q_{b^{*}}^{*}), \\
      \left.\frac{\partial^{2}  LR(q_{i}, q_{b^{*}}^{*})}{\partial q_{i}q_{b^{*}}}\right|_{(q_{i},q_{b^{*}})=(q_{i}^{*},q_{b^{*}}^{*})}&=\left.\frac{\partial  LR(q_{i}, q_{b^{*}}^{*})}{\partial q_{i}}\right|_{q_{i}=q_{i}^{*}} \left.\frac{\partial  LR(q_{i}^{*}, q_{b^{*}})}{\partial q_{b^{*}}}\right|_{q_{b^{*}}=q_{b^{*}}^{*}}\frac{1}{ LR(q_{i}^{*}, q_{b^{*}}^{*})}.
\end{align*}
Hence, all we need to show is the identification result of $ LR(q_{i}^{*}, q_{b^{*}}^{*})$, $ \partial  LR(q_{i}^{*}, q_{b^{*}}^{*})/\partial q_{i}$ and $\partial  LR(q_{i}^{*}, q_{b^{*}}^{*})/\partial q_{b^{*}}$. By definition of $Q_{i}(\mathbf{Z})$, it holds from Assumption \ref{new_assumption_imp} that
\begin{align*}
    \left.\frac{\partial Q_{i}(\mathbf{Z})}{\partial W_{i}}\right|_{\mathbf{Z}=\mathbf{Z}^{*}}&=f_{\tilde{U}_{i,k}}(\tilde{R}_{i,k}(\mathbf{Z}^{*}))\left.\frac{\partial \tilde{R}_{i,k}(\mathbf{Z})}{\partial W_{i}}\right|_{\mathbf{Z}=\mathbf{Z}^{*}} \\
    &=f_{\tilde{U}_{i,k}}(F^{-1}_{\tilde{U}_{i,k}}(q_{i}^{*}))\alpha, \\
      \left.\frac{\partial^{2} Q_{j}(\mathbf{Z})}{\partial W_{i}^{2}}\right|_{\mathbf{Z}=\mathbf{Z}^{*}}
    &=\frac{\partial }{\partial W_{i}}\left(f_{\tilde{U}_{i,k}}(\tilde{R}_{i,k}(\mathbf{Z}^{*}))\left.\frac{\partial \tilde{R}_{i,k}(\mathbf{Z})}{\partial W_{i}}\right|_{\mathbf{Z}=\mathbf{Z}^{*}}\right), \\
        &=\frac{\partial f_{\tilde{U}_{i,k}}(F_{\tilde{U}_{i,k}}^{-1}(q_{i}^{*}))}{\partial \tilde{u}_{i,k}}\left(\left.\frac{\partial \tilde{R}_{i,k}(\mathbf{Z})}{\partial W_{i}}\right|_{\mathbf{Z}=\mathbf{Z}^{*}}\right)^{2}+f_{\tilde{U}_{i,k}}(\tilde{R}_{i,k}(\mathbf{Z}^{*}))\left.\frac{\partial^{2}\tilde{R}_{i,k}(\mathbf{Z})}{\partial W_{i}^{2}}\right|_{\mathbf{Z}=\mathbf{Z}^{*}} \\
        &=\frac{\partial f_{\tilde{U}_{i,k}}(F_{\tilde{U}_{i,k}}^{-1}(q_{i}^{*}))}{\partial \tilde{u}_{i,k}}\alpha^{2}.
\end{align*}
Hence, we have
\begin{align*} 
LR(q_{i}^{*}, q_{b^{*}}^{*})&=\frac{\left.\left.\partial Q_{i}(\mathbf{Z})\right/\partial W_{i}\right|_{\mathbf{Z}=\mathbf{Z}^{*}}}{\left.\left.\partial Q_{b^{*}}(\mathbf{Z})\right/\partial W_{b^{*}}\right|_{\mathbf{Z}=\mathbf{Z}^{*}}} \\ 
\left.\frac{\partial  LR(q_{i}, q_{b^{*}}^{*})}{\partial q_{i}}\right|_{q_{i}=q_{i}^{*}}&=\frac{\left.\left.\partial^{2} Q_{i}(\mathbf{Z})\right/\partial W_{i}^{2}\right|_{\mathbf{Z}=\mathbf{Z}^{*}}}{\left(\left.\left.\partial Q_{i}(\mathbf{Z})\right/\partial W_{i}\right|_{\mathbf{Z}=\mathbf{Z}^{*}}\right)^{2}}\frac{\left.\left.\partial Q_{i}(\mathbf{Z})\right/\partial W_{i}\right|_{\mathbf{Z}=\mathbf{Z}^{*}}}{\left.\left.\partial Q_{b^{*}}(\mathbf{Z})\right/\partial W_{b^{*}}\right|_{\mathbf{Z}=\mathbf{Z}^{*}}}, \\
\left.\frac{\partial  LR(q_{i}^{*}, q_{b^{*}})}{\partial q_{b^{*}}}\right|_{q_{b^{*}}=q_{b^{*}}^{*}}&=-\frac{\left.\left.\partial^{2} Q_{b^{*}}(\mathbf{Z})\right/\partial W_{b^{*}}^{2}\right|_{\mathbf{Z}=\mathbf{Z}^{*}}}{\left(\left.\left.\partial Q_{b^{*}}(\mathbf{Z})\right/\partial W_{b^{*}}\right|_{\mathbf{Z}=\mathbf{Z}^{*}}\right)^{2}}\frac{\left.\left.\partial Q_{i}(\mathbf{Z})\right/\partial W_{i}\right|_{\mathbf{Z}=\mathbf{Z}^{*}}}{\left.\left.\partial Q_{b^{*}}(\mathbf{Z})\right/\partial W_{b^{*}}\right|_{\mathbf{Z}=\mathbf{Z}^{*}}}.
\end{align*}
Therefore, the required result holds.

\end{proof}

\end{document}